\documentclass[aps,prb,twocolumn,showpacs,10pt,floatfix]{revtex4-2}

\usepackage{natbib}
\usepackage{amsfonts}
\usepackage{amsmath}
\usepackage{amssymb}
\usepackage{braket}
\usepackage{mathtools}
\usepackage{rotating}
\usepackage{color}
\usepackage{framed}
\usepackage{enumerate}
\usepackage{graphicx}
\usepackage{float}
\usepackage{hyperref}
\hypersetup{
    colorlinks=true,
    linkcolor=blue,    
    urlcolor=blue,
    citecolor=blue
    }
\usepackage{tikz}
\usepackage{circuitikz}
\usepackage{tabularx}
\usepackage{multirow}
\usepackage{csquotes}

\definecolor{blue-plot}{rgb}{0, 0.4, 0.8}
\definecolor{orange-plot}{rgb}{1, 0.6, 0}

\definecolor{blue-1}{rgb}{0.776,0.832,0.872}
\definecolor{blue-2}{rgb}{0.746,0.802,0.842}
\definecolor{blue-3}{rgb}{0.806,0.862,0.902}
\definecolor{gray-1}{rgb}{0.88,0.892,0.896}
\definecolor{gray-2}{rgb}{0.85,0.862,0.866}
\definecolor{gray-3}{rgb}{0.91,0.922,0.926}

\newcommand{\vm}{\vec m}
\newcommand{\vE}{\vec E}

\newcommand{\vi}{\vec i}

\renewcommand{\vr}{\vec r}
\newcommand{\vis}{{\vec i}_{\rm s}}
\newcommand{\ve}{\vec e}
\newcommand{\vmu}{\mbox{\boldmath $\mu$}}
\newcommand{\vu}{{\vec u}}
\newcommand{\vus}{{\vec u}_{\rm s}}
\newcommand{\vnabla}{\mbox{\boldmath $\nabla$}}

\newcommand{\vmus}{\vmu_{\rm s}}
\renewcommand{\vec}[1]{\mathbf{#1}}
\newcommand{\Dex}{D_{\rm ex}}
\newcommand{\zzeta}{f}
\newcommand{\calzzeta}{{\cal F}}
\newcommand{\tildew}{\varphi}
\newcommand{\tilder}{\rho}

\begin{document}
\title{Theory of nonlinear magnetoelectric transport effects in normal-metal -- magnetic-insulator heterostructures}

\author{Oliver Franke}
\author{Piet W. Brouwer}
\affiliation{Dahlem Center for Complex Quantum Systems and Physics Department, Freie Universit\"at Berlin, Arnimallee 14, 14195 Berlin, Germany}

\begin{abstract}
Heterostructures of normal metals (N) and magnetic insulators (F) show paradigmatic effects, such as spin-Hall magnetoresistance and electric drag currents. These effects are linear in the applied electric field $E(\omega)$. Normal-metal--magnetic-insulator heterostructures also exhibit a characteristic nonlinear response quadratic in $E(\omega)$, referred to as unidirectional spin-Hall magnetoresistance or spin-torque diode effect. In this article, we develop a theory of the bilinear response of FN bilayers and NFN trilayers for finite frequencies $\omega$ of the driving field and for four contributions that have been previously considered in the literature: Joule heating, phonon-mediated unidirectional magnetoresistance, the spin-torque diode effect, and magnonic unidirectional spin-Hall magnetoresistance. We identify their distinct dependencies on frequency and the magnetization direction of the magnetic insulator and examine their scaling with magnetic field and system geometry, providing a framework for experimental differentiation.
\end{abstract}

\maketitle

\newpage

\section{Introduction}
\label{sec:introduction}

The combination of non-magnetic metals and magnetically ordered materials is at the heart of spintronics. Their interfaces couple spin, charge, and heat transport and give rise to key spintronic effects driven by spin accumulations, electric fields, and temperature gradients \cite{Hirohata2020-jp}. In this article, we consider bilayer and trilayer systems consisting of normal metals and a magnetic insulator. Such multilayers exhibit a rich set of spintronic phenomena, which arise from the conversion of electronic excitations in the normal metal to magnonic excitations in the ferromagnetic insulator at their interface. Prominent such spintronic effects are current-induced magnetization switching \cite{Avci2017-kj,Yang2024-fm}, spin-Hall magnetoresistance \cite{Weiler2012-gh,Huang2012-fu,Nakayama2013-gf,Hahn2013-rw,Vlietstra2013-uv,Althammer2013-zm,Lotze2014-qv,Choi2017-fn,Chen2013-gf,Chen2016-pc,Zhang2019-zv}, and the spin-Seebeck and spin-Peltier effects \cite{Uchida2010-je, Bauer2012-tx, Flipse2014-kh}.

The spin-Hall magnetoresistance refers to a dependence of the in-plane conductivity of a bilayer consisting of a normal metal (N) and a ferromagnetic insulator (F) on the magnetization direction of F \cite{Weiler2012-gh,Huang2012-fu,Nakayama2013-gf,Hahn2013-rw,Vlietstra2013-uv,Althammer2013-zm,Lotze2014-qv,Choi2017-fn,Chen2013-gf,Chen2016-pc,Zhang2019-zv}. Central to this effect is that the spin-Hall effect (SHE) in N drives a spin current between N and F, which, via the inverse spin-Hall effect (ISHE), has a back action on the in-plane charge current in N. Since the magnitude of the spin current between N and F depends on the magnetization direction in F, the resulting change of the in-plane conductivity of N depends on it, too.
Magnetotransport effects have also been observed in N$|$F$|$N trilayers, in which application of an electric field in one N layer leads to a current response in the other. This nonlocal response is known as magnon-mediated current drag \cite{Zhang2012-fy,Zhang2012-ig,Kajiwara2010-rj,Cornelissen2015-fh,Goennenwein2015-lb,Schlitz2021-ho,Li2016-ye,Wu2016-cs,Muduli2020-ub}.

Being a linear-response effect, the spin-Hall magnetoresistance is invariant under reversal of either the current direction or the magnetization direction. These symmetries are broken in the current response quadratic in the applied electric field $E$, which is therefore referred to as {\em unidirectional spin-Hall magnetoresistance} (USMR) \cite{Avci2015-pk,Avci2018-uv,Zhang2016-si}. Unidirectional magnetotransport effects have been proposed as a building block for spintronic applications, such as multi-state memory devices \cite{Avci2017-xh}. The USMR effect was originally observed in multilayers involving a metallic ferromagnet, but subsequently predicted \cite{Wang2018-qd,Sterk2019-mt} and measured \cite{Liu2021-ny} in bilayers featuring a ferromagnetic insulator.

\begin{figure*}
    \centering
    \includegraphics[width=1\textwidth]{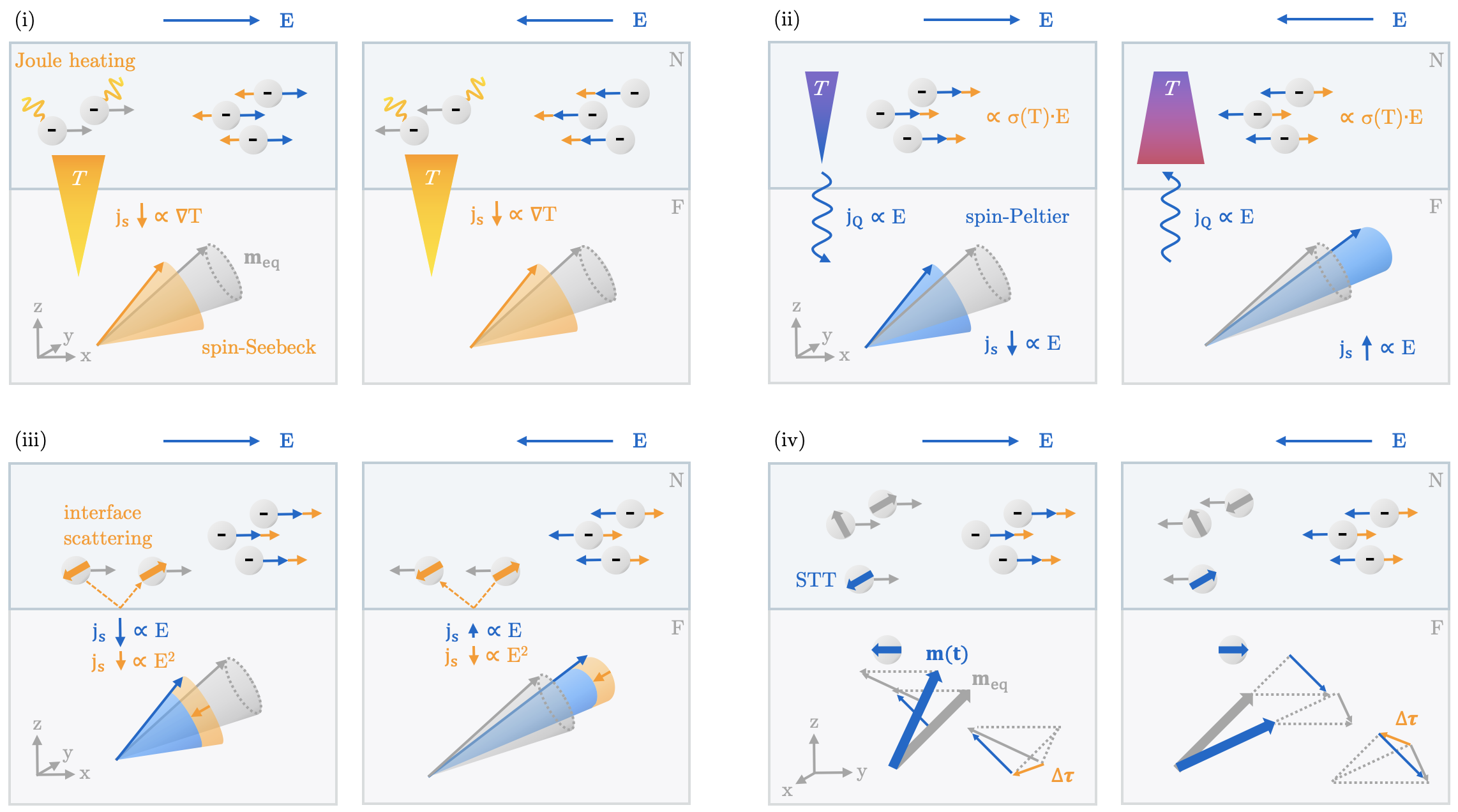}
    \caption{Schematic illustration of the four sources of unidirectional spin-Hall magnetoresistance in ferromagnetic-insulator--normal-metal multilayers considered here: (i) Joule heating in N drives a spin current from N to F via the spin-Seebeck effect. This spin current affects the in-plane charge current in N via the inverse spin-Hall effect (ISHE). (ii) Via the spin-Peltier effect, the linear-in-$E$ spin current from N to F associated with the spin-Hall magnetoresistance effect changes the temperature in N. The temperature change implies a conductivity change and, hence, a quadratic-in-$E$ contribution to the in-plane current response. (iii) Stimulated magnon excitation from spin-flip scattering at the F$|$N interface is intrinsically nonlinear. This gives a nonlinear dependence of the spin current through the F$|$N interface on the applied field $E$ and, via the ISHE, of the charge response in N. (iv) Analogous to the spin-torque diode effect, a strong coherent excitation of magnetization modes in F causes a modulation of all linear response coefficients and, hence, a unidirectional response. The coordinate system of F in panel (iv) is rotated with respect to N and other panels for clarity. Processes linear in the electric field $E$ are colored in blue, while processes quadratic in $E$ are orange.}
    \label{fig:1}
\end{figure*}

In the literature, different sources of a nonlinear spin-Hall magnetotransport have been proposed for multilayers containing a ferromagnetic insulator, see Fig.\ \ref{fig:1}: (i) Joule heating in N, in combination with the spin-Seebeck \cite{Uchida2010-je, Bauer2012-tx} and ISHE effects \cite{Cornelissen2015-fh}, (ii) a linear-in-$E$ conductivity change of N  \cite{Sullivan2023-mw} from the heating or cooling of N by the spin-Peltier effect \cite{Flipse2014-kh}, (iii) the intrinsic nonlinearity of stimulated magnon emission from spin-flip scattering at the F$|$N interface \cite{Sterk2019-mt}, and (iv) a modulation of the linear response from the coherent driving of large-amplitude magnetization modes in F \cite{Chiba2014-mh,Liu2011-sy,Kondou2012-kt,Ganguly2014-zw,Schreier2015-pt,Sklenar2015-gm}, analogous to the spin-torque diode effect \cite{Tulapurkar2005-ew, Sankey2006-cf}. Since the conductivity of N is dominated by electron-phonon scattering, the second mechanism is referred to as ``phonon-mediated USMR''. Since these four mechanisms for the USMR effect have almost identical nonlinear signatures in electrical transport, distinguishing them requires an understanding of their dependence of relevant system parameters, such as the full dependence on magnetization direction or the sample geometry.

In this article, we calculate the USMR effect for these four sources of the nonlinear response --- which for brevity we refer to as ``Joule-heating'', ``phonon-mediated'', ``interfacial'', and ``spin-torque'' USMR --- in a common theoretical framework, so that we can compare their magnitude and characteristic dependence on the magnetization direction. Motivated by recent experiments that address spintronic effects on ultrafast time scales \cite{Fulop2020-do,Kampfrath2013-ke,Schellekens2014-se,Kimling2017-qd}, we calculate these contributions to the USMR effect for driving frequencies $\omega$ up to the THz frequency range. In addition, we consider unidirectional magnon-mediated current drag --- the nonlocal counterpart of the USMR effect --- in a trilayer geometry \cite{Cornelissen2015-fh}, where the driving field $E$ and the current response are in different layers. The frequency dependence and the comparison between local and nonlocal responses give additional possibilities to distinguish the different mechanisms for USMR. We find that, depending on the driving frequency $\omega$, of the four mechanisms mentioned above, all but the interfacial USMR mechanism can dominate the quadratic-in-$E$ charge current response.

Our work builds on and unifies a large number of previous theoretical works in this field. This includes geometric considerations of the magnetization direction dependence of the USMR effect \cite{Liu2021-ny, Avci2015-pk}, magnetization dynamics in F based on the Landau-Lifshitz-Gilbert equation, which is either solved numerically \cite{Liu2021-ny, Wang2018-qd} or analytically for a uniformly precessing mode \cite{Chiba2014-mh, Ganguly2014-zw, Chiba2015-jt}, diffusion equations for spin and heat transport in N and F \cite{Sullivan2023-mw, Sterk2019-mt, Chiba2015-jt, Wang2018-qd}, and nonperturbative relations between spin accumulations and temperatures across F$|$N interfaces \cite{Sterk2019-mt}.
Our theory of the nonlinear spin-Hall magnetoresistance strongly relies on theories of the linear spin-Hall magnetoresistance effect. For F$|$N bilayers, such theories were put forward in Refs.\ \onlinecite{Chen2013-gf,Chen2016-pc,Zhang2019-zv} in the zero-frequency limit and in Ref.\ \onlinecite{Reiss2021-em} for finite frequencies. In our companion article \cite{Franke2025-lin} we extend the theory of Ref.\ \onlinecite{Reiss2021-em} to nonlocal linear response in an N$|$F$|$N trilayer, accounting for the driving of coherent magnetization modes via the spin-Hall effect as well as for diffusive transport of thermal magnons in F \cite{Cornelissen2016-wy}. In the present article, we use the fundamental linear-response relations derived in Ref.\ \onlinecite{Franke2025-lin} as the starting point for our calculations. 

One aspect in which the present article goes beyond the existing literature on the USMR effect is that we consider driving frequencies $\omega$ up to the THz regime. In this frequency range, the frequency dependence of nonlinear effects in N$|$F$|$N trilayers arises primarily from magnon transport and the coherent excitation of magnetization modes in F. Both effects are fully captured within the linear response theory of the spin-Hall magnetoresistance effect \cite{Reiss2021-em,Franke2025-lin}. Relevant time scales are the diffusion time for thermal magnons across F \cite{Cornelissen2016-wy,Schmidt2021-qo} and the inverse spectral widths of coherent magnetization modes in F \cite{Chiba2014-mh, Ganguly2014-zw, Chiba2015-jt}. The former time scales correspond to characteristic frequencies in the THz range, making the nonlinear response of such systems highly frequency-dependent in this regime. In contrast, the linear transport and sources of nonlinearity in N and at the F$|$N interface have only a weak intrinsic frequency dependence for $\omega$ in the THz range and below \cite{Schellekens2014-se,Razdolski2017-st,Kampfrath2013-ke,Seifert2018-vj}.

The remainder of this article is organized as follows: In Sec.~\ref{sec:system}, we introduce the system of interest and the bilinear conductivity, which describes the quadratic-in-$E$ response to an applied electric field $E$. In Sec.~\ref{sec:linearresponse}, we review the spin-dependent transport equations for the normal-metal layers and briefly discuss the structure of the linear-response theory of Ref.\ \onlinecite{Franke2025-lin}, as far as necessary for the calculation of the quadratic-in-$E$ response considered here.
The theory of the quadratic-in-$E$ current response is developed in Sec.~\ref{sec:nonlinearresponse}. We present our result in terms of four response coefficients, calculated separately for each of the four sources of nonlinear response we consider, which each represent a different characteristic magnetization-direction dependence of the bilinear conductivity. Numerical estimates for material and device parameters of a typical Pt$|$YIG$|$Pt trilayer are given in Sec.~\ref{sec:numericalestimates}. We conclude in Sec.~\ref{sec:conclusion}. The appendices contain additional details of our calculations and a brief summary of the linear-response results of Ref.\ \onlinecite{Franke2025-lin} used for our calculations. To facilitate the evaluation of our results for device parameters not considered by us or for other material combinations, an open source code is available to evaluate local and nonlocal, linear and nonlinear response for different materials and system sizes \cite{Franke2025-zenodo}.

\section{N$|$F$|$N Trilayer and Bilinear Response}
\label{sec:system}

\begin{figure}
    \centering
    \includegraphics[width=0.4\textwidth]{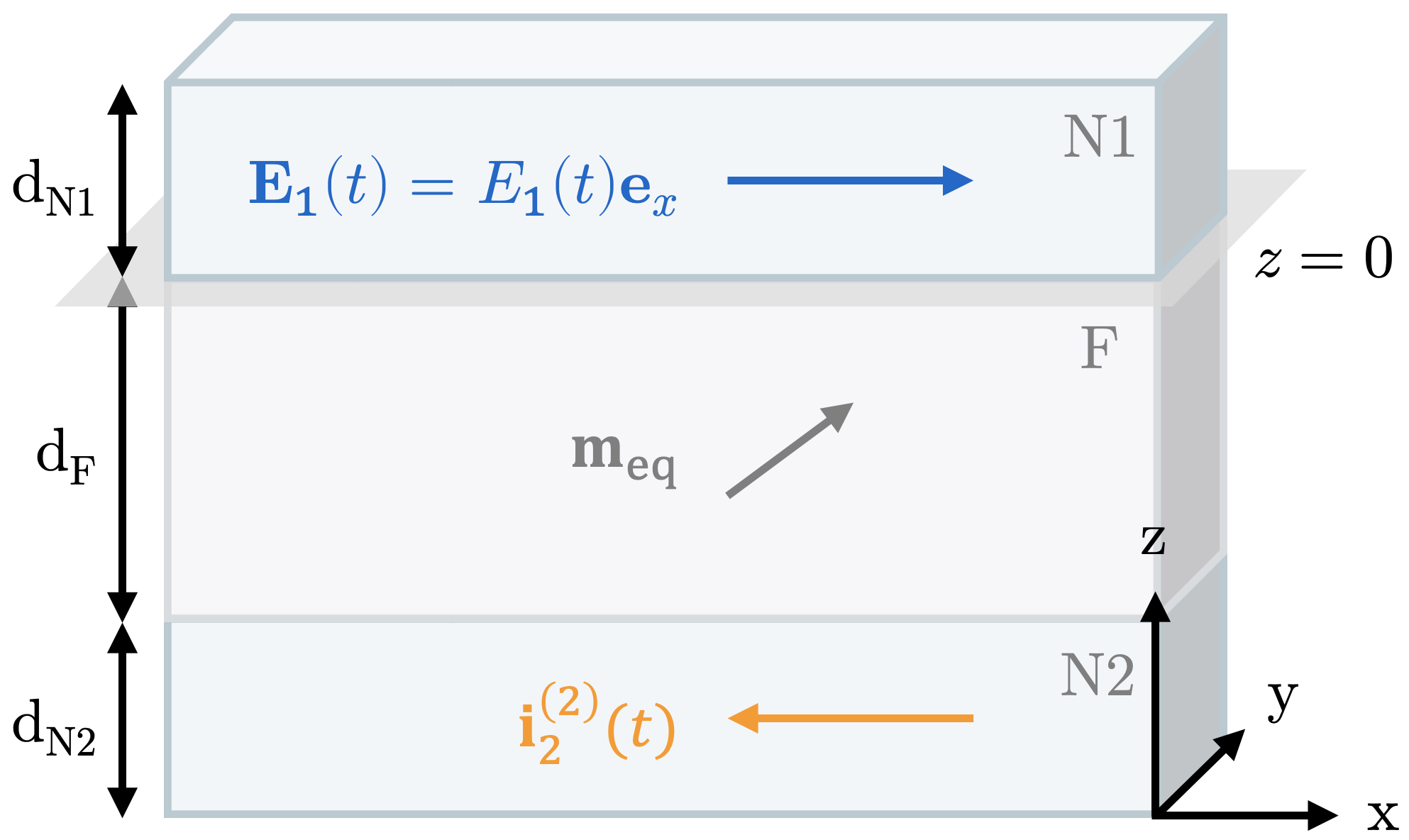}
\caption{Geometry of the N$|$F$|$N trilayer, consisting of two normal metals N1 and N2 and a magnetically ordered insulator F. An in-plane electric field $\vE_j(t) = E_j(t) \ve_x$ in one of the normal-metal layers gives rise to charge currents in both layers. In this article, we calculate the current contribution $\vi_{i}^{(2)}$ quadratic in the applied field $E_j$ for driving fields $E_j(t) \propto \cos(\omega t)$ with driving frequencies ranging from the {\it dc} limit $\omega = 0$ to the THz regime.}
\label{fig:geometry}
\end{figure}

We consider an N$|$F$|$N trilayer consisting of two normal metals N1 and N2, separated by a ferromagnetic insulator F, see Fig.\ \ref{fig:geometry}. We choose coordinates such that N1 and N2 are located at $0 < z < d_{{\rm N}1}$ and $-d_{\rm F} - d_{{\rm N}2} < z < -d_{\rm F}$, whereas the ferromagnetic insulator F is at $-d_{\rm F} < z < 0$. 
The coupling strength between F and N$i$, $i=1,2$, is set by the spin mixing conductance $g_{\uparrow\downarrow i}$ of that interface \cite{Brataas2000-ar,Xia2002-xc}. If desired, results for an N$|$F bilayer can be obtained from the charge response of the N$|$F$|$N trilayer we consider here by setting $g_{\uparrow\downarrow 2} = 0$.

The ferromagnetic insulator has a magnetization direction indicated by the unit vector
\begin{equation}
    \mathbf{m}_{\rm eq} = m_x \mathbf{e}_x + m_y \mathbf{e}_y + m_z \mathbf{e}_z.
\label{eq:equilibriummagnetization}
\end{equation}
To describe the direction perpendicular to $\vm_{\rm eq}$, we choose a complex unit vector $\mathbf{e}_\perp$, whose real and imaginary parts span the perpendicular plane to $\vm_{\rm eq}$ and that fulfills
\begin{equation}
    \mathbf{e}_\perp \times \vm_{\rm eq} = i \mathbf{e}_\perp.
\label{eq:basiscrossproduct}
\end{equation}
Using the complex basis $(\vm_{\rm eq},\ve_{\perp},\ve_{\perp}^*)$, the magnetization $\vm(z,t)$ can then be decomposed into longitudinal and transverse components as
\begin{equation}
  \vm(z,t) = m_{\parallel}(z,t) \vm_{\rm eq}
  + m_{\perp}(z,t) \ve_{\perp} + m_{\perp}^*(z,t) \ve_{\perp}^*,
  \label{eq:magbasism}
\end{equation}
where $m_{\perp}(z,t)$ is complex and $m_{\parallel}^2 = 1 - 2 |m_{\perp}|^2$.

Spatially uniform time-dependent electric fields $E_j(t) \ve_x$, $j=1,2$, are applied in N1 and N2, with
\begin{equation}
  E_j(t) = \frac{1}{2 \pi} \int_{-\infty}^{+\infty} \mathrm{d}\omega E_j(\omega) e^{-i \omega t}.
\label{eq:fourier}
\end{equation}
Up to quadratic order in the applied fields, the spatially averaged current response may be written as
\begin{equation}
  \bar i_i^{x/y}(t) = \bar i_i^{x/y}(t)^{(1)} + \bar i^{x/y}_i(t)^{(2)},
  \label{eq:charge_current_definition}
\end{equation}
where $\bar i_i^{x,y}(t)^{(1)}$ is linear in the applied electric field $E_j$ and $\bar i^{x/y}_i(t)^{(2)}$ is bilinear, {\em i.e.} it is proportional to a sum of products $E_j E_k$ with $j$, $k=1,2$.
The linear response $\bar i_i^{x/y}(t)^{(1)}$ is considered in detail in our companion article \cite{Franke2025-lin}.
Fourier transforming to time, the bilinear response $\bar i^{x/y}_i(t)^{(2)}$ can be expressed as
\begin{align}
  \bar i^{x}_{i}(\Omega)^{(2)} =&\,
  \sum_{j,k=1}^{2} \int \frac{{\rm d}\omega}{2 \pi}
  \sigma^{xxx(2)}_{ijk}(\omega_+,\omega_-) E_j(\omega_+) E_k(\omega_-)^*,
  \nonumber \\
  \bar i^{y}_{i}(\Omega)^{(2)} =&\,
  \sum_{j,k=1}^{2} \int \frac{{\rm d}\omega}{2 \pi}
  \sigma^{yxx(2)}_{ijk}(\omega_+,\omega_-) E_j(\omega_+) E_k(\omega_-)^*,
  \label{eq:bilinear}
\end{align}
where $\omega_{\pm} = \omega \pm \Omega/2$. In this article, we calculate the bilinear conductivity $\sigma^{x/yxx(2)}_{ijk}(\omega_+,\omega_-)$ for the four sources of nonlinear response discussed in the introduction.
Since the calculation of the bilinear response builds on the calculation of linear response, we first briefly review the currents that flow in linear response to the applied electric field, before we turn to the bilinear response in Sec.\ \ref{sec:nonlinearresponse}.

\section{Linear response}
\label{sec:linearresponse}

The calculation of the quadratic-in-applied-field response builds on the linear-response theory. A complete calculation of the local and nonlocal linear current response to a time-dependent electric field and/or a time-dependent Joule heating term is given in our companion article \cite{Franke2025-lin}. We here review the main results of that article, as far as they are necessary for a theory of the quadratic response, referring to Ref.\ \onlinecite{Franke2025-lin} and App.\ \ref{sec:impedances} for details.

{\em Charge and spin transport.---}
Transport of charge and spin in the normal metals N1 and N2 is coupled via the spin-Hall effect and the inverse spin-Hall effect. To linear order in the applied electric field $E^x_i(\omega)$ and the gradients of the induced spin accumulation $\vu_{\rm s}(z,\omega)$, the charge current densities $i^{x,y}(z,\omega)$ satisfy the phenomenological response equations \cite{Dyakonov1971-pp,Dyakonov1971-yh,Hirsch1999-gc,Takahashi2006-tz}
\begin{align}
  i^x (z, \omega) =&\ \sigma_{{\rm N}i} E_i(\omega) - \theta_{\text{SH}i} \frac{\sigma_{{\rm N}i}}{2} \frac{\partial}{\partial z} u_{{\rm s}y} (z, \omega) , \label{eq:ix} \\
  i^y (z, \omega) =&\ \theta_{\text{SH}i} \frac{\sigma_{{\rm N}i}}{2} \frac{\partial}{\partial z} u_{{\rm s}x} (z, \omega), \label{eq:iy}
\end{align}
where $\theta_{{\rm N}i}$, $\sigma_{{\rm N}i}$, and $\vu_{{\rm s}}(z,\omega)$ are the spin-Hall angle, the linear electrical conductivity, and the spin accumulation in N$i$, respectively. (The indices $i=1,2$ are written in accordance with the position $z$ in N1 or N2.) Here, and in the following, we denote spatial directions by superscripts and spin directions by subscripts or boldface vector notation. The spin accumulation is defined as $e u_{{\rm s}z} = \mu_{\uparrow} - \mu_{\downarrow}$, where $\mu_{\sigma}$ is the chemical potential for electrons of spin $\sigma$, the spin direction being defined with respect to the $z$-axis, with analogous definitions for $u_{{\rm s}x}$ and $u_{{\rm s}y}$.
Upon integrating Eqs.\ (\ref{eq:ix}) and (\ref{eq:iy}) to $z$, the spatially averaged corrections $\delta \bar i_{i}^{x,y}(\omega)$ to the charge current densities in N1 and N2 can be expressed in terms of the spin accumulations at the interface $\vu_{{\rm s}i}(\omega)$ \cite{Chen2013-gf,Chen2016-pc},
\begin{align}
  \label{eq:deltaix}
  \delta \bar i_{i}^{x}(\omega) =&\ (-1)^{i-1} \theta_{{\rm SH}i}
     \frac{  \sigma_{{\rm N}i}}{2 d_{{\rm N}i}} 
     u_{{\rm s}iy}(\omega), \\
  \label{eq:deltaiy}
  \delta \bar i_{i}^{y}(\omega) =&\ -(-1)^{i-1} \theta_{{\rm SH}i}
     \frac{\sigma_{{\rm N}i}}{2 d_{{\rm N}i}} 
    u_{{\rm s}ix}(\omega),\ \ i=1,2.
\end{align}
Because the dynamical variables in N1, N2, and F are mainly needed at the ferromagnet--normal-metal interfaces at $z=0$ and $z=-d_{\rm F}$, we use the short-hand notation $\vu_{{\rm s}1}(t) = \vus(0,t)$, $\vu_{{\rm s}2}(t) = \vus(-d_{\rm F},t)$ and analogously for the other variables.
(The exception to this notation is the charge current density $\bar i_{j}^{x,y}(t)$ of Eq.~\eqref{eq:charge_current_definition}, which is the average over the cross section of layer $j$.)

To find the spin accumulations at the interfaces $\vu_{{\rm s}i}(\omega)$, we combine the phenomenological response equations for the spin current density $\vis^z(z,\omega)$ in N$i$ \cite{Dyakonov1971-pp,Dyakonov1971-yh,Hirsch1999-gc,Takahashi2006-tz},
\begin{align}
  \vis^z(z, \omega) =&\ - \frac{\sigma_{{\rm N}i}}{2} \frac{\partial}{\partial z} \vus (z, \omega) - \theta_{\text{SH}i} \sigma_{{\rm N}i} E_i(\omega) \ve_{y},
  \label{eq:jsz}
\end{align}
with the continuity equation for the spin density
\begin{equation}
  -i e^2 \omega \nu_{{\rm N}i} \vus(z,\omega)
  + \frac{\partial}{\partial z} \vis^z(z,\omega)
  = - e^2 \frac{\nu_{{\rm N}i}}{\tau_{{\rm sf},i}}
  \vu_{{\rm s}}(z,\omega).
  \label{eq:spincontinuity}
\end{equation}  
Here $\nu_{{\rm N}i}$ and $\tau_{{\rm sf},i}$ are the electronic density of states and the spin-flip time in N$i$, $i=1,2$, respectively. The spin current density is defined as $i^z_{{\rm s}z} = i_{\uparrow}^z - i_{\downarrow}^z$, where $i^z_{\sigma}$ is the charge current carried with spin $\sigma$, defined with respect to the $z$-axis, with analogous definitions for $i^z_{{\rm s}x}$ and $i^z_{{\rm s}y}$. Solving Eqs.\ (\ref{eq:jsz}) and (\ref{eq:spincontinuity}) one finds a relation between the spin accumulation $\vu_{{\rm s}i}(\omega)$ and spin current $\vi_{{\rm s}i}(\omega)$ at the interface, which we write in the concise form \cite{Reiss2021-em}
\begin{equation}
  \label{eq:nmspinrelation}
  (-1)^{i-1} Z_{{\rm N}i}(\omega) \vi^z_{{\rm s}i}(\omega) =
  \vu_{{\rm s}i}(\omega) - \delta \vu_{{\rm s}i}(\omega),
\end{equation}
where $Z_{{\rm N}i}(\omega)$ is the ``spin impedance'' of N$i$,
\begin{equation}
  Z_{{\rm N}i}(\omega) = \frac{2 \lambda_{{\rm N}i}(\omega)}{\sigma_{{\rm N}i}}
\label{eq:zn}
\end{equation}
with the spin relaxation length
\begin{equation}
  \lambda_{{\rm N}i}(\omega)^2 = \frac{\sigma_{{\rm N}i}}{2 e^2 \nu_{{\rm N}i}(1/\tau_{{\rm sf},i} - i \omega)},
  \label{eq:lambdaNi}
\end{equation}
and $\delta \vu_{{\rm s}i}(\omega)$ a source term proportional to the applied electric field,
\begin{equation}
  \delta \vu_{{\rm s}i}(\omega) = 2 (-1)^{i-1}
  \lambda_{{\rm N}i}(\omega) \theta_{{\rm SH}i} E_i \ve_y.
\label{eq:linearsourceterm}
\end{equation}
Since we will only consider frequencies $\omega \ll 1/\tau_{{\rm sf},i}$, we neglect the frequency dependence of $\lambda_{{\rm N}i}$ --- and consequentially $Z_{{\rm N}i}$.

{\em Heat transport.---} Since the F$|$N interface couples spin and heat transport, the linear response to the applied electric field also includes a change $\Delta T_{\rm e}(z,\omega)$ of the temperature away from its equilibrium value $T$. To relate $\Delta T_{\rm e}(z,\omega)$ and the heat current $j_{\rm eQ}^z(z,\omega)$ in N$i$ to their values at the F$|$N interfaces, we combine the phenomenological equation for thermal conductivity, 
\begin{equation}
  j^z_{{\rm Q}}(z,\omega) = - \kappa_{{\rm e}i} \frac{\partial}{\partial z} \Delta T_{\rm e}(z,\omega),
  \label{eq:heatkappa}
\end{equation}
with the continuity equation for heat transport,
\begin{align}
  -i \omega C_{{\rm e}i} \Delta T_{{\rm e}i}(z,\omega)
  + \frac{\partial}{\partial z} j^z_{{\rm }Qi}(z,\omega)
  =&\ - \frac{C_{{\rm e}i}}{\tau_{{\rm ep},i}} 
  \Delta T_{{\rm e}i}(z,\omega) \nonumber \\ &\ \mbox{}
  + s(z,\omega),
\label{eq:heatcontinuity}
\end{align}
where $\kappa_{{\rm e}i}$, $C_{{\rm e}i}$, and $\tau_{{\rm ep},i}$ are the electronic contribution to the thermal conductivity and the heat capacity and the characteristic electron-phonon relaxation time in N$i$, $i=1,2$, respectively, and $s(z,\omega)$ is a source term from Joule heating. To bring about the formal analogy to charge and spin transport, we measure $j^z_{\rm Q}$ and $\Delta T_{\rm e}$ in units of an equivalent charge current density $i^z_{\rm eQ}$ and voltage $u_{\rm eQ}$,
\begin{align}
  i_{{\rm Q}}^z(z,t) =&\, \frac{2e}{k_{\rm B} T} j^z_{{\rm Q}}(z,t), 
    \label{eq:equivalentcurrents} \\
  u_{{\rm e}Q}(z,t) &= \frac{k_{\rm B}}{e} \Delta T_{\rm e}(z,t).
    \label{eq:equivalentvoltages}
\end{align}
Solving Eqs.\ (\ref{eq:heatkappa}) and (\ref{eq:heatcontinuity}) with the boundary condition that the heat current vanishes at the interfaces with vacuum at $z = d_{{\rm N}1}$ and $z = -d_{{\rm N}2} - d_{{\rm F}}$, we find that the relation between the temperature change and heat currents at the normal-metal--ferromagnet interfaces can be cast in a form identical to that of Eq.~\eqref{eq:nmspinrelation},
\begin{equation}
  \label{eq:nmheatrelation}
  (-1)^{i-1} Z_{{\rm QN}i}(\omega) i^z_{{\rm Q}i}(\omega) =
  u_{{\rm eQ}i}(\omega) - \delta u_{{\rm eQ}i}(\omega).
\end{equation}
Here, the ``thermal impedance'' is
\begin{equation}
  \label{eq:zqn}
  Z_{{\rm QN}i}(\omega) = \frac{k_{\rm B}^2 T}{2 e^2}
  \frac{l_{{\rm ep},i}(\omega)}{\kappa_{{\rm e}i}}
  \coth \frac{d_{{\rm N}i}}{l_{{\rm ep},i}(\omega)}
\end{equation}
with the thermal relaxation length
\begin{align}
  l_{{\rm ep},i}(\omega)^2 &= \frac{\kappa_{{\rm e}i} \tau_{{\rm ep},i}}{C_{{\rm e}i}(1 -i \omega\tau_{{\rm ep},i})}, \ \ i=1,2,
  \label{eq:thermalrelaxationlength}
\end{align}
whereas
\begin{align}
 \label{eq:nonlin_deltauqe}
 \delta u_{{\rm eQ}1}(\omega)
  =&\,
  \frac{k_{\rm B} l_{{\rm ep},1}(\omega)}{e \kappa_{{\rm e}1}}
  \int_0^{d_{{\rm N}1}} dz'
  \frac{\cosh \frac{d_{{\rm N}1}-z'}{l_{{\rm ep},1}(\omega)}}{\sinh\frac{d_{{\rm N}1}}{l_{{\rm ep},1}(\omega)}}
  s_1(z',\omega) 
\end{align}
is a source term representing the effect of Joule heating in N$1$, with an analogous expression for $\delta u_{{\rm eQ}2}(\omega)$. 

{\em Longitudinal and transverse spin transport.---}
In the description of the response of the full N$|$F$|$N trilayer, we find it useful to decompose vector-valued variables, such as the spin accumulations $\vu_{{\rm s}i}(t)$, the spin currents $\vi_{{\rm s}i}(t)$, or the source terms $\delta \vu_{{\rm s}i}(t)$ into components parallel to and perpendicular to the magnetization direction $\vm_{\rm eq}$. The parallel components describe spin-flip scattering at the F$|$N interfaces and spin transport by thermal magnons in F \cite{Bender2015-tr,Bender2012-ll,Cornelissen2016-wy, Reiss2022-tm}; the transverse components couple to coherent magnetization dynamics \cite{Tserkovnyak2002-ax,Tserkovnyak2002-hn}. Using the complex basis $(\vm_{\rm eq}, \ve_{\perp}, \ve_{\perp}^*)$, see Eq.\ (\ref{eq:magbasism}), we 
write
\begin{align}
  \vu_{{\rm s}i}(t) =&\, u_{{\rm s}i \parallel}(t) \vm_{\rm eq}
  + u_{{\rm s}i \perp }(t) \ve_{\perp} + u_{{\rm s}i \perp }^*(t) \ve_{\perp}^*,
  \label{eq:magbasis}
\end{align}
with analogous expressions for $\vi_{{\rm s}i}(t)$ and $\delta \vu_{{\rm s}i}(t)$. The relation (\ref{eq:nmspinrelation}) then applies to longitudinal ($\parallel$) and transverse ($\perp$) components separately.
Note that the time-domain variables $i_{{\rm s}\perp}(z,t)$, $u_{{\rm s}\perp}(z,t)$, and $\delta u_{{\rm s}\perp}(z,t)$ are complex, so that their Fourier transforms at frequencies $\omega$ and $-\omega$ are not complex conjugates of each other.

At the F$|$N interfaces and inside F, the heat current $i^z_{{\rm Q}i}(t)$ is coupled to the longitudinal component $i_{{\rm s}i\parallel}^z(t)$ \cite{Cornelissen2016-wy, Reiss2022-tm}. To simplify expressions for coupled spin and heat transport, we therefore combine $i_{{\rm s}i\parallel}^z(t)$ and $i^z_{{\rm Q}i}(t)$ into a single two-component vector,
\begin{equation}
  {\cal I}_{i}(\omega) = \begin{pmatrix} i_{{\rm s}i\parallel}^z(\omega) 
  \\ i^z_{{\rm Q}i}(\omega) \end{pmatrix},
  \label{eq:calI}
\end{equation}
with similar definitions for the {\em generalized spin accumulation} ${\cal U}_{{\rm e}i}(\omega) = (u_{{\rm s}i \parallel}(\omega),u_{{\rm eQ}i}(\omega))^{\rm T}$ and $\delta {\cal U}_{{\rm e}i}(\omega) = (\delta u_{{\rm s}i \parallel}(\omega),\delta u_{{\rm eQ}i}(\omega))^{\rm T}$. 
We also combine the impedances $Z_{{\rm N}i}$ and $Z_{{\rm QN}i}$ for spin and heat transport into a $2 \times 2$ matrix
\begin{equation}
  {\cal Z}_{{\rm N}i}(\omega) = \begin{pmatrix} Z_{{\rm N}i} & 0 \\ 0 & Z_{{\rm QN}i}(\omega) \end{pmatrix}.
\label{eq:znmat}
\end{equation}

{\em Interface with ferromagnet.---} To complete the linear-response theory, equations governing spin and heat transport across the two normal-metal--ferromagnet interfaces at $z = 0$ and $z = -d_{\rm F}$ and inside the ferromagnetic insulator for $-d_{\rm F} < z < 0$ are needed. The equations governing spin and heat transport across the two F$|$N interfaces will be discussed in Sec.\ \ref{sec:nonlinearresponse}, where we also consider the leading nonlinear corrections to these equations. For the equations governing spin and heat transport inside the ferromagnetic insulator F we refer to Ref.~\onlinecite{Reiss2021-em} and the companion article \cite{Franke2025-lin}. 

The solution of all linear-response equations, including those not shown here explicitly, can be summarized in terms of separate linear relations between the source term $\delta u_{{\rm s}j \perp}(\omega)$ and the transverse spin accumulation $u_{{\rm s}j \perp}(\omega)$ and between the two-component source term $\delta {\cal U}_{{\rm e}i}(\omega)$ and the generalized spin accumulation ${\cal U}_{{\rm e}i}(\omega)$,
\begin{align}
\begin{split}
  u_{{\rm s}i \perp}(\omega) =&\, \sum_{j=1}^{2} \zzeta_{ij\perp}(\omega)
  \delta u_{{\rm s}j \perp}(\omega), \\
  {\cal U}_{{\rm e}i}(\omega) =&\, \sum_{j=1}^{2} \calzzeta_{ij}(\omega)
  \delta {\cal U}_{{\rm e}j}(\omega),
  \end{split}
\label{eq:linresponse}
\end{align}
where the dimensionless coefficients $\zzeta_{ij\perp}(\omega)$ and $\calzzeta_{ij}(\omega) = \calzzeta_{ij}(-\omega)^*$ are complex numbers and $2 \times 2$ matrices, respectively. 
Explicit expressions for $\zzeta_{ij\perp}(\omega)$ and $\calzzeta_{ij}(\omega)$ in terms of the spin mixing conductances $g_{\uparrow\downarrow i}$ of the N$|$F interfaces and properties of F are given in Ref.~\onlinecite{Franke2025-lin} and App.~\ref{sec:impedances}. The current densities $\bar i^{x,y}_{i}(\omega)$ follow immediately upon substitution of Eq.~\eqref{eq:linresponse} into Eqs.~\eqref{eq:deltaix} and \eqref{eq:deltaiy}.

In the next Section, we repeatedly apply Eq.\ (\ref{eq:linresponse}) at the frequencies $\Omega$, $\omega_+$ and $\omega_-$, with suitably chosen source terms $\delta u_{{\rm s}i\perp}(\omega)$ and $\delta {\cal U}_{{\rm e}i}(\omega)$, to find the bilinear conductivities $\sigma^{x/yxx(2)}_{ijk}(\omega_+,\omega_-)$. The frequency dependence of the bilinear response is dominated by the frequency dependence of the dimensionless linear-response coefficients $\zzeta_{ij\perp}(\omega)$ and $\calzzeta_{ij}(\omega)$. As is described in detail in Ref.\ \onlinecite{Franke2025-lin}, the longitudinal response coefficient $\calzzeta_{ij}(\omega)$ is a smooth function of $\omega$, whereas the transverse response coefficient $\zzeta_{ij\perp}(\omega)$ has sharp maxima in the vicinity of the frequencies $\omega_n$ of resonant magnetization modes of F.

\section{Bilinear response}
\label{sec:nonlinearresponse}
In this Section, we calculate the nonlinear conductivities $\sigma^{x/yxx(2)}_{ijk}(\omega_+,\omega_-)$ for the ``Joule-heating'', ``phonon-mediated'', ``interfacial'', and ``spin-torque'' mechanisms discussed in the introduction.
For these four mechanisms we find that the dependence of $\sigma^{(2)}_{ijk}(\omega_+,\omega_-)$ on the magnetization direction $\vm_{\rm eq}$ is of the form
\begin{widetext}
\begin{align}
    \sigma^{xxx(2)}_{ijk}(\omega_+,\omega_-) &= \frac{\sigma_{{\rm N}i}}{d_{{\rm N}i}} 
    \left[ v_{ijk}(\omega_+,\omega_-) m_y + r_{ijk}(\omega_+,\omega_-) m_y(1-m_y^2) \right], 
    \label{eq:sigmaxxx} \\
    \sigma^{yxx(2)}_{ijk}(\omega_+,\omega_-) &= - \frac{\sigma_{{\rm N}i}}{d_{{\rm N}i}} 
    \left[ w_{ijk}(\omega_+,\omega_-) m_x + r_{ijk}(\omega_+,\omega_-) m_x(1-m_y^2) + t_{ijk}(\omega_+,\omega_-) m_y m_z \right],
    \label{eq:sigmayxx}
\end{align}
\end{widetext}
where $v_{ijk}(\omega_+,\omega_-)$, $w_{ijk}(\omega_+,\omega_-)$, $r_{ijk}(\omega_+,\omega_-)$, and $t_{ijk}(\omega_+,\omega_-)$ are response coefficients with the dimension of $[\mbox{length}]/[\mbox{electric field}]$.
The bilinear response of Eqs.~\eqref{eq:sigmaxxx} and \eqref{eq:sigmayxx} changes sign under a $\pi$ rotation of the magnetization direction $\vm_{\rm eq}$ around the $z$-axis. For $\vm_{\rm eq}$ in the $xy$ plane this is consistent with the magnetization-direction dependence of the phenomenological theory of Ref.~\onlinecite{Avci2015-pk} and with the experimental observations in Refs.~\onlinecite{Liu2021-ny} and \onlinecite{Han2020-gl}. (No out-of-plane magnetization directions were considered in these references.) Equations \eqref{eq:sigmaxxx} and \eqref{eq:sigmayxx} also imply that there is no bilinear response in the $x$ direction, {\em i.e.}, in the direction of the applied electric field, if $\vm_{\rm eq}$ is in the $xz$-plane, whereas the bilinear response in the $y$-direction disappears if $\vm_{\rm eq}$ is in the $y$-direction.

In the following four Subsections, we calculate the response coefficients $v_{ijk}(\omega_+,\omega_-)$, $w_{ijk}(\omega_+,\omega_-)$, $r_{ijk}(\omega_+,\omega_-)$, and $t_{ijk}(\omega_+,\omega_-)$ for each source of bilinear response separately.
In Sec.~\ref{sec:numericalestimates} we then numerically evaluate these results using parameter values for a Pt$|$YIG$|$Pt trilayer. 
Additional details of the calculation can be found in App.~\ref{app:bilinear}.

\subsection{Joule heating contribution}
\label{sec:joule_nonlin}

The Joule heating rate $s(z,t)$, see Eq.~(\ref{eq:heatcontinuity}), has contributions from charge and spin currents \cite{Tulapurkar2011-lm, Taniguchi2016-ei}, which for our geometry read 
\begin{align}
  s(z,t) =&\,
  E_i(t) i^x(z,t) - \frac{1}{2} \frac{\partial}{\partial z}
  (\vus(z,t) \cdot \vi_{{\rm s}}^z(z,t)).
  \label{eq:jouleheatingrate}
\end{align}
Here we neglect corrections to the Joule heating from the spin-Hall effect and its inverse since they are proportional to the square of the spin-Hall angle and approximate
\begin{equation}
  s_i(t) = \sigma_{{\rm N}i} E_i(t)^2
  \label{eq:jouleheatingrateresponse}
\end{equation}
for the normal layer N$i$, $i=1,2$. Complete expressions, which include the Joule heating from dissipative spin currents, are given in App.\ \ref{app:bilinear}.

Inserting the approximation (\ref{eq:jouleheatingrateresponse}) into Eq.~\eqref{eq:nonlin_deltauqe} and performing a Fourier transform, the source term $\delta u_{{\rm eQ}i}(\Omega)$ in Eq.~\eqref{eq:nmheatrelation} becomes
\begin{align}
  \delta u_{{\rm eQ}j}(\Omega) =&\,
  \frac{k_{\rm B} l_{{\rm ep},j}(\Omega)^2}{2 \pi e \kappa_{{\rm e}i}}
  \sigma_{{\rm N}j} \int d\omega
  E_j(\omega_+) E_j(\omega_-)^*,
\label{eq:nonlin_deltauqeresponse}
\end{align}
where $\omega_{\pm} = \omega \pm \Omega/2$.

According to Eq.\ (\ref{eq:linresponse}), the source term $\delta u_{{\rm eQ}i}(\Omega)$ gives rise to a longitudinal spin accumulation $u_{{\rm s}\parallel j}(\Omega)$ at frequency $\Omega$.
The quadratic-in-$E$ charge current densities can then be calculated from the inverse spin-Hall effect, see Eqs.\ (\ref{eq:deltaix}) and (\ref{eq:deltaiy}).
We find that the charge current associated with Joule heating is of the form (\ref{eq:bilinear}), where the non-zero response coefficients $v_{ijk}$ and $w_{ijk}$ defined in Eqs.~(\ref{eq:sigmaxxx}) and (\ref{eq:sigmayxx}) read
\begin{align}
  v_{ijk}^{\rm Jo}(\omega_+,\omega_-) &= w_{ijk}^{\rm Jo}(\omega_+,\omega_-) 
  \label{eq:joule}
  \\
  &= (-1)^{i-1} \theta_{{\rm SH}i}
  \calzzeta_{ij}(\Omega)_{12}
  \frac{k_{\rm B} l_{{\rm ep},j}(\Omega)^2}{2 e \kappa_{{\rm e}j}}
  \sigma_{{\rm N}j} \delta_{jk}, \nonumber
\end{align}
where the $2 \times 2$ matrix $\calzzeta_{ij}(\Omega)$ was introduced in Eq.\ (\ref{eq:linresponse}).
The coefficient $t_{ijk}(\omega_+,\omega_-)$ vanishes for the Joule heating mechanism. The coefficient $r_{ijk}(\omega_+,\omega_-)$ vanishes for the approximation (\ref{eq:jouleheatingrateresponse}), but not if the contribution from dissipative spin currents is included. We refer to App.\ \ref{app:bilinear} for the complete expressions.

\subsection{Phonon-mediated contribution}
\label{sec:phononumr}

Via the spin-Peltier effect, a spin accumulation at the F$|$N interface causes a heat current through the interface and, hence, a change $\Delta T_{{\rm e}i}$ of the electron temperature in N1 and N2 that is linear in the applied electric field. The temperature change at the F$|$N interfaces is
\begin{align}
  \Delta T_{{\rm e}i}(\omega) =&\,
  \frac{2 e}{k_{\rm B}} m_y \sum_{j} (-1)^{j-1} \calzzeta_{ij}(\omega)_{21} 
  \lambda_{{\rm N}j} \theta_{{\rm SH}j} E_j(\omega),
\end{align}
see Eqs.\ (\ref{eq:linearsourceterm}), (\ref{eq:equivalentvoltages}), and (\ref{eq:linresponse}). Away from the interface, one has
\begin{equation}
  \Delta T_{\rm e}(z,\omega) =
  \Delta T_{{\rm e}1}(\omega) \frac{\cosh[(z-d_{{\rm N}1})/l_{{\rm ep},1}]}{\cosh(d_{{\rm N}1}/l_{{\rm ep},1})}
\end{equation}
for N1, see Eqs.\ (\ref{eq:heatkappa}) and (\ref{eq:heatcontinuity}), with an analogous expression for N2. The temperature change $\Delta T_{\rm e}(z,\omega)$ comes with a change in conductivity, which, together with the applied field itself, leads to a current response quadratic in $E$. Since the conductivity of the normal layers is dominated by electron-phonon scattering, this contribution to the bilinear response essentially involves coupling to the phonon bath, which is why it is referred to as ``phonon-mediated'' USMR \cite{Sullivan2023-mw}.

So far we have neglected any change of the phonon temperature and treated the phonons as a bath at fixed temperature $T$. This approximation is justified because the phonon heat capacity $C_{\rm p}$ is typically much larger than the electronic heat capacity. For a description of phonon-mediated USMR, we need to take the change of the phonon temperature into account. Neglecting the phonon thermal conductivity, one has
\begin{equation}
  \Delta T_{\rm p}(z,\omega) = g_{i}(\omega) \Delta T_{{\rm e}}(z,\omega),\ \
  i=1,2,
\end{equation}
with
\begin{equation}
  g_{i}(\omega)^{-1} = 1 + \left(1- i \omega \tau_{{\rm p},i} \right) \frac{C_{{\rm p}i}}{C_{{\rm e}i}} \frac{\tau_{{\rm ep},i}}{\tau_{{\rm p},i}}, \label{eq:gi}
\end{equation}
where $\tau_{{\rm p},i}$ is the relaxation time for heat conduction to the substrate.
(Equation (\ref{eq:heatcontinuity}) for the electron temperature does not change upon taking into account the change of the phonon temperature in the limit $C_{{\rm e}i} \ll C_{{\rm p}i}$, see App.\ \ref{app:bilinear}.)

The temperature dependence of the conductivity is
\begin{equation}
  \sigma_{{\rm N}}(z,t) = \sigma_{{\rm N}i} (1 - \alpha_{Ti} \Delta T_{{\rm p}}(z,t)),
\label{eq:sigman_tcr}
\end{equation}
with $\alpha_{Ti}$ the temperature coefficient of resistance of N$i$. There is a direct bilinear contribution to the charge current $\bar i^x_i(\Omega)$, which follows from the first term in Eq.\ (\ref{eq:ix}). This direct contribution has the form \eqref{eq:bilinear} with
\begin{align}
   \label{eq:phonon}
 v_{ijk}^{\rm ph}(\omega_+,\omega_-) =&\, - 
  \delta_{ik} (-1)^{j-1} \theta_{{\rm SH}j}\calzzeta_{ij}(\omega_+)_{21}
    \lambda_{{\rm N}j}
   \\ \nonumber &\, \mbox{} \times
  \frac{2e}{k_{\rm B}} \alpha_{Ti} g_i(\omega_+) 
  l_{{\rm ep},i}(\omega_+) 
  \tanh \frac{d_{{\rm N}i}}{l_{{\rm ep},i}(\omega_+)}.
\end{align}
The coefficients $w_{ijk}^{\rm ph}$, $r_{ijk}^{\rm ph}$, and $t_{ijk}^{\rm ph}$, which determine the Hall response, vanish for this direct phonon-mediated bilinear contribution to the charge current \cite{Sullivan2023-mw}. In App.\ \ref{app:bilinear} we also consider a small, indirect contribution to the charge current, which follows from the effect that $\Delta T_{\rm e}(z,\omega)$ has on the spin accumulation in N1 and N2. For this indirect effect, all response coefficients are nonzero.

\citet{Sullivan2023-mw} measured and calculated this contribution to the USMR for driving frequency up to $10^5 \, \mathrm{Hz}$. Their theoretical model relies on diffusion equations for spin and temperature in F and N and coupling via the spin-mixing conductance at the F$|$N interface. In our approach, this part of the calculation is absorbed in the linear response matrix $\calzzeta_{ij}$, see our companion article \cite{Franke2025-lin}.

\subsection{Interfacial contribution}
\label{sec:interface_nonlin}
The transport of spin and heat via incoherent magnons through ferromagnet--normal-metal interfaces is governed by the spin-mixing conductance $g_{\uparrow\downarrow i}$ \cite{Brataas2000-ar,Xia2002-xc,Xiao2010-wi,Bender2015-tr}. The longitudinal component of the spin current $i_{{\rm s}\parallel i}$ and the heat currents $i_{{\rm Q}i}$ through the F$|$N$i$ interface, which together form the two-component vector ${\cal I}_i$, see Eq.\ (\ref{eq:calI}), depend on the (generalized) spin accumulation ${\cal U}_{{\rm e}i} = (u_{{\rm s}i \parallel},u_{{\rm eQ}i})^{\rm T}$ in N$i$ and its counterpart ${\cal U}_{{\rm m}i} = (-\mu_{{\rm m}i}/e, k_{\rm B} \Delta T_{{\rm m}i}/e)^{\rm T}$ in F, where $\mu_{{\rm m}i}$ and $\Delta T_{{\rm m}i}$ are the magnon chemical potential and excess temperature at the interface with N$i$, $i=1,2$.
To leading order in $g_{\uparrow\downarrow i}$, but without restriction to small potential differences, the longitudinal spin and energy current densities through the F$|$N$i$ interface read \cite{Bender2015-tr,Bender2012-ll,Cornelissen2016-wy, Reiss2022-tm}
\begin{align}
  {\cal I}_{i} =&\
  \frac{8 e \mbox{Re}\, g_{\uparrow\downarrow i}}{\hbar^2 s} 
  (-1)^{i-1}
  \int d\varepsilon \nu_{\rm m}(\varepsilon)
  (\varepsilon - e u_{{\rm s}i\parallel})
  \begin{pmatrix}
  1 \\ \frac{2 \varepsilon}{k_{\rm B} T} 
  \end{pmatrix}
  \nonumber \\ &\, \times
  \left[ f\left(\frac{\varepsilon - e u_{{\rm m}i}}{k_{\rm B} T_{\rm m}} \right)
  - f\left(\frac{\varepsilon - e u_{{\rm s}i}}{k_{\rm B} T_{\rm e}} \right)
  \right],
  \label{eq:ispin}
\end{align}
where $f(z) = 1/(e^z-1)$ the Planck function, $s$ is the spin per volume in F, $\nu_{\rm m}(\varepsilon)$ the magnon density of states, and we use the two-component vector notation of Eq.\ (\ref{eq:calI}).
We assume a quadratic magnon dispersion $\varepsilon(k) = \hbar(\omega_0 + \Dex k^2)$, with $\omega_0$ the ferromagnetic resonance frequency and $\Dex$ the spin stiffness, which gives
\begin{equation}
  \nu_{\rm m}(\varepsilon) = \frac{1}{4 \pi^2 \hbar \Dex} \sqrt{\frac{\varepsilon - \hbar \omega_0}{\hbar \Dex}}.
  \label{eq:num}
\end{equation}

Expanding Eq.~\eqref{eq:ispin} to second order in the differences $u_{{\rm s}i\parallel}-u_{{\rm m}i}$ and $u_{{\rm eQ}i}-u_{{\rm mQ}i} = k_{\rm B} (\Delta T_{{\rm e}i} - \Delta T_{{\rm m}i})/e$ and taking the limit $\hbar \omega_0/k_{\rm B} T \to 0$ gives \cite{Cornelissen2016-wy, Reiss2021-em}
\begin{align}
  {\cal U}_{{\rm e}i} - {\cal U}_{{\rm m}i}
  =&\, - (-1)^{i-1} {\cal Z}_{{\rm FN}i\parallel}
  \left( {\cal I}_{i} + \delta {\cal I}_{i}^{\rm in} \right),
\label{eq:fnrelationpara}
\end{align}
where $\mathcal{Z}_{{\rm FN}i \parallel}$ is the $2 \times 2$ interfacial impedance matrix,
\begin{equation}
    \mathcal{Z}_{{\rm FN}i}^{-1} = \frac{3 k_{\rm T}^3 \mathrm{Re} g_{\uparrow\downarrow i}}{16 \pi^{3/2} s} 
    \begin{pmatrix}
        4 \zeta(3/2) & 10 \zeta(5/2) \\
        10 \zeta(5/2) & 35 \zeta(7/2)
    \end{pmatrix},
\label{eq:calzfnpara}
\end{equation}
with $\zeta$ the Riemann zeta function and $k_{\rm T}$ the thermal magnon wave number,
\begin{equation}
  k_{\rm T} = \sqrt{\frac{k_{\rm B} T}{\hbar D_{\rm ex}}}.
\label{eq:thermalk}
\end{equation}
The matrix impedance ${\cal Z}_{{\rm FN}i}$ describes the coupled spin and heat transport through the F$|$N interface \cite{Cornelissen2016-wy}, which includes the linear longitudinal spin impedance \cite{Bender2015-tr} and the interfacial spin-Seebeck effect \cite{Xiao2010-wi}. The correction $\delta {\cal I}_i$ is quadratic in the current density ${\cal I}_{i}$ and the generalized spin accumulation ${\cal U}_{{\rm e}i}$ \cite{Sterk2019-mt},
\begin{align}
  \label{eq:nonlin_interfacesources}
  \delta {\cal I}_{i\alpha}^{\rm in}
  =&\,
  \sum_{\beta,\gamma=1}^{2}
  \left[
  (-1)^{i-1} 
  {\cal U}_{{\rm e}i\beta} {\cal A}_{\alpha\beta\gamma}
  + {\cal I}_{i\beta} {\cal B}_{i,\alpha\beta\gamma} \right]
  {\cal I}_{i\gamma}
\end{align}
where explicit expressions for the rank-three tensors ${\cal A}_i$ and ${\cal B}_i$ are given in App.\ \ref{app:bilinear}. The Greek indices refer to the two-component matrix notation of Eq.\ (\ref{eq:calI}). 

For a calculation of the bilinear charge current response, the linear-response equations must be solved with the source term $\delta {\cal I}_i(\Omega)$ in Eq.\ (\ref{eq:fnrelationpara}), but without the sources $\delta u_{{\rm s}i\perp}(\Omega)$ and $\delta {\cal U}_{{\rm e}i}$ in Eqs.\ (\ref{eq:nmspinrelation}) and (\ref{eq:nmheatrelation}). 
This is most easily accomplished if we change variables to the generalized spin accumulation $\tilde {\cal U}_{{\rm e}i}(\Omega)$,
\begin{equation}
  \label{eq:tildeu_substitution}
  \tilde {\cal U}_{{\rm e}i}(\Omega) =
  {\cal U}_{{\rm e}i}(\Omega) +
  (-1)^{i-1} {\cal Z}_{{\rm FN}i \parallel}
  \delta {\cal I}_{i}^{\rm in},
\end{equation}
which, by construction, obeys Eq.\ (\ref{eq:fnrelationpara}) without the source term $\delta {\cal I}^{\rm in}_i$ and Eq.\ (\ref{eq:nmspinrelation}) and (\ref{eq:nmheatrelation}) with sources $\delta \tilde u_{{\rm s}i\perp}(\Omega) = 0$, $(\delta \tilde u_{{\rm s}i\parallel}(\Omega), \delta \tilde u_{{\rm eQ}i}(\Omega))^{\rm T} = (-1)^{i-1} {\cal Z}_{{\rm FN}i \parallel} \delta {\cal I}_{i}^{\rm in}$. Since these are the same source terms as in the original linear-response problem, the linear-response relation (\ref{eq:linresponse}) can be used to calculate $\tilde {\cal U}_{{\rm e}i}(\Omega)$. From there, we find the spin accumulation $u_{{\rm s}i\parallel}(\Omega)$ from Eq.~\eqref{eq:tildeu_substitution}. For the bilinear charge current response we thus find $v_{ijk}^{\rm in} = w_{ijk}^{\rm in} = -r_{ijk}^{\rm in}$ with
\begin{align}
  \lefteqn{v_{ijk}^{\rm in}(\omega_+,\omega_-) =
  2 \sum_{l=1}^{2}  (-1)^{i+j+k+l}
  \theta_{{\rm SH}i} \theta_{{\rm SH}j} \theta_{{\rm SH}k} 
  \lambda_{{\rm N}j}\lambda_{{\rm N}k}} ~~
  \nonumber \\ 
  &\, \mbox{} \times \! \! \! \! \!
  \sum_{\alpha,\beta,\gamma = 1}^{2}
  [\tilde \calzzeta_{il}(\Omega)
  {\cal Z}_{{\rm FN}l\parallel}]_{1 \alpha}
  [{\cal Z}_{{\rm N}l}(\omega_-)^{*-1} 
  \tilde \calzzeta_{lk}(\omega_-)^*]_{\gamma 1}
  \label{eq:interface} \\ 
  &\,  \mbox{} \times
  \{
  [{\cal Z}_{{\rm N}l}(\omega_+)^{-1} 
  \tilde \calzzeta_{lj}(\omega_+)]_{\beta 1}
  {\cal B}_{l,\alpha\beta\gamma}
  - \calzzeta_{lj}(\omega_+)_{\beta 1}
  {\cal A}_{\alpha\beta\gamma} \}, \nonumber
\end{align}
where we abbreviated 
\begin{equation}
  \tilde \calzzeta_{ij}(\omega) = \delta_{ij}\openone - \calzzeta_{ij}(\omega).
\end{equation}
The response coefficient $t^{\rm in}_{ijk}(\omega_+,\omega_-) = 0$.

Sterk {\em et al.} \cite{Sterk2019-mt} also use Eq.\ (\ref{eq:ispin}) as the starting point of their theoretical analysis of the interface contribution to the USMR. Our calculation differs from that of Sterk {\em et al.} in that we account for the coupled spin and heat transport across the interface, whereas Ref.\ \onlinecite{Sterk2019-mt} only considers the spin current $i^z_{{\rm s}i\parallel}$ in response to a difference of spin accumulation/magnon potential across the interface, {\em i.e.}, restricts to the 11-component of Eq.\ (\ref{eq:nonlin_interfacesources}). 

\subsection{Spin-torque contribution}
\label{sec:precession_nonlin}

Since the physical mechanisms for spin transport parallel and perpendicular to the magnetization direction are different, there are separate equations governing longitudinal and transverse spin transport through the F$|$N interfaces. The fundamental equation for longitudinal spin transport through the F$|$N interfaces is Eq.\ (\ref{eq:fnrelationpara}), which describes longitudinal spin transport in combination with heat transport. Here, we are interested in transport linear in the (generalized) spin accumulation difference ${\cal U}_{{\rm e}i} - {\cal U}_{{\rm m}i}$, which allows us to neglect the source term $\delta {\cal I}_i^{\rm in}(t)$ in Eq.\ (\ref{eq:fnrelationpara}).

The fundamental equation for the  transverse linear spin transport through the interface between F and N is \cite{Tserkovnyak2002-ax,Tserkovnyak2002-hn}
\begin{align}
  u_{{\rm s}i \perp}(t) + i \frac{\hbar}{e} \dot m_{\perp i}(t)
  =&\, -(-1)^{i-1}
  \frac{i_{{\rm s}i\perp}(t)}{g_{\uparrow\downarrow i}}.
  \label{eq:ff2}
\end{align}
Here, $m_{\perp i}(t)$ is the complex transverse magnetization component at the F$|$N$i$ interface, see Eq.\ (\ref{eq:magbasism}).

The longitudinal and transverse components, as they appear in Eqs.\ (\ref{eq:fnrelationpara}) and (\ref{eq:ff2}), are taken with respect to the instantaneous magnetization direction $\vm_i(t)$. For linear response, the difference between $\vm_i(t)$ and the equilibrium magnetization direction $\vm_{\rm eq}$ may be neglected and one works with longitudinal and transverse components defined with respect to $\vm_{\rm eq}$, as was done in Sec.\ \ref{sec:linearresponse}. Taking into account the difference between $\vm_i(t)$ and $\vm_{\rm eq}$ to leading order in $m_{\perp i}$ gives a quadratic-in-$E$ correction to the current response. We refer to this contribution to the nonlinear response as the ``spin-torque'' contribution, because it derives from the spin-torque driven coherent magnetization dynamics of F \cite{Tulapurkar2005-ew, Taniguchi2016-ei}. A theory for this effect was previously formulated by Chiba, Bauer, and Takahashi \cite{Chiba2014-mh} for an F$|$N bilayer in the macrospin approximation and without accounting for longitudinal spin currents and heat transport across the interface. \citet{gems2025-fr} cast the theory of Ref.\ \onlinecite{Chiba2014-mh} in the magneto-electric circuit formulation, which took account of spin relaxation by longitudinal spin currents across the F$|$N interface, and included coherent magnetization modes beyond the uniform one. The full theory of the spin-torque diode contribution must also contain a contribution to the nonlinear response from spin currents by thermal magnons, which was not included in Refs.\ \onlinecite{Chiba2014-mh,gems2025-fr}.

In App.\ \ref{app:bilinear} we show that shifting from longitudinal/transverse components defined with respect to $\vm_i(t)$ to components defined with respect to $\vm_{\rm eq}$ amounts to the inclusion of source terms into Eqs.\ (\ref{eq:fnrelationpara}) and (\ref{eq:ff2}), so that they now read
\begin{align}
  {\cal U}_{{\rm e}i}(t) - {\cal U}_{{\rm m}i}(t) =&\, - (-1)^{i-1} {\cal Z}_{{\rm FN}\parallel} [{\cal I}_i(t) + \delta {\cal I}^{\rm to}_i(t)], \nonumber \\
  u_{{\rm s}i \perp}(t) + i \frac{\hbar}{e} \dot m_{\perp i}(t)
  =&\, -(-1)^{i-1} 
  \frac{i_{{\rm s}i\perp}(t) + \delta i_{{\rm s}i\perp}^{\rm to}(t)}{g_{\uparrow\downarrow i}},\!
  \label{eq:ff22}
\end{align}
where
\begin{align}
  \label{eq:jsource}
  \delta {\cal I}^{\rm to}_{i\alpha}(t)
  =&\, - 2 \mbox{Re}\, m_{\perp i}^*(t)
  i_{{\rm s}i\perp}(t) \delta_{\alpha,1}
  \\ &\, \mbox{} -2 (-1)^{i-1} \mbox{Re}\, m_{\perp i}^*(t)
  u_{{\rm s}\perp i}(t)
  \left( {\cal Z}_{{\rm FN}i\parallel}^{-1} \right)_{\alpha 1}, \nonumber \\
   \delta i_{{\rm s}i\perp}^{\rm to}(t) =&\ m_{\perp i}(t) i_{{\rm s}i\parallel}(t) + (-1)^{i-1} g_{\uparrow\downarrow i} m_{\perp i}(t) u_{{\rm s}i\parallel}(t). \nonumber 
\end{align}
The first term of the longitudinal source current $\delta {\cal I}^{\rm to}_{i\alpha}$ is associated with the coherent magnetization dynamics. This is the source of nonlinear response that was considered in Ref.~\onlinecite{gems2025-fr}. The second term contributing to $\delta {\cal I}^{\rm to}_{i\alpha}$ and the first term contributing to $\delta i_{{\rm s}i\perp}^{\rm to}$ involve spin currents carried by thermal magnons and were not considered in Refs.\ \onlinecite{Chiba2014-mh,gems2025-fr}.

The source terms $\delta {\cal I}_i^{\rm to}$ and $\delta i_{{\rm s}i\perp}^{\rm to}$ may be calculated from the linear-response theory of Sec.~\ref{sec:linearresponse}. The magnetization amplitude $m_{\perp i}(\omega)$ can be calculated from Eqs.\ (\ref{eq:nmspinrelation}), (\ref{eq:linresponse}), and (\ref{eq:ff2}), from which we find
\begin{align}
  m_{\perp i}(\omega) = -
  \sum_{j=1}^{2} (-1)^{j-1}  \eta_{{ij\perp}}(\omega)
  \theta_{{\rm SH}j} \sigma_{{\rm N}j} 
  E_j(\omega) \ve_{\perp}^* \cdot \ve_y,
  \label{eq:mperpresponse}
\end{align}
where 
\begin{align*}
  \eta_{ij\perp}(\omega) =&\
  \frac{e}{\hbar \omega} \frac{Z_{{\rm N}j}}{Z_{{\rm N}i}}
  \left(Z_{{\rm N}j} \delta_{ij} - (Z_{{\rm N}i} + g_{\uparrow\downarrow i}^{-1}) \tilde \zzeta_{ij\perp}(\omega)
  \right)
\end{align*}
and we abbreviated 
\begin{equation}
  \tilde \zzeta_{ij\perp}(\omega) = \delta_{ij} - \zzeta_{ij\perp}(\omega).
\end{equation}
We then calculate the charge current response by solving the linear-response equations with the additional source terms $\delta i_{{\rm s}i\perp}(\Omega)$ and $\delta {\cal I}(\Omega)$ of Eqs.\ (\ref{eq:fnn2}) and (\ref{eq:fnn1}), but without the source terms $\delta u_{{\rm s}i\perp}(\Omega)$, $\delta u_{{\rm s}i\parallel}(\Omega)$, and $\delta u_{{\rm eQ}i}(\Omega)$ in Eqs.\ (\ref{eq:nmspinrelation}) and (\ref{eq:nmheatrelation}). Proceeding as in Sec.\ \ref{sec:interface_nonlin} we find that a quadratic-in-$E$ contribution to the charge current is given by response coefficients
\begin{align}
    w_{ijk}^{\rm to}(\omega_+,\omega_-) =&\ \tildew_{ijk}(\omega_+,\omega_-) + \tildew_{ijk}(-\omega_+,-\omega_-)^*, \nonumber \\
    r_{ijk}^{\rm to}(\omega_+,\omega_-) =&\ \tilder_{ijk}(\omega_+,\omega_-) + \tilder_{ijk}(-\omega_+,-\omega_-)^*  \nonumber \\ &\, - \tildew_{ijk}(\omega_+,\omega_-) 
  - \tildew_{ijk}(-\omega_+,-\omega_-)^* , \nonumber \\
    t_{ijk}^{\rm to}(\omega_+,\omega_-) =&\ i \tildew_{ijk}(\omega_+,\omega_-) - i \tildew_{ijk}(-\omega_+,-\omega_-)^*. \nonumber 
\label{eq:torque}
\end{align}
where we abbreviated
\begin{align*}
\begin{split}
     \tildew_{ijk}(\omega_+,\omega_-) =&\ (-1)^{i+j+k-1} \theta_{{\rm SH}i} \theta_{{\rm SH}j} \theta_{{\rm SH}k} \lambda_{{\rm N}k} \sigma_{{\rm N}j} \\
     &\, \times \sum_{l=1}^{2} \tilde \zzeta_{il\perp}(\Omega) \big[ \eta_{lj\perp}(\omega_+) \calzzeta_{lk}(\omega_-)^* \\
     &\, - g_{\uparrow\downarrow l}^{-1} \eta_{lj\perp}(\omega_+) \mathcal{Z}^{-1}_{{\rm N}l}(\omega_-)^* \tilde \calzzeta_{lk}(\omega_-)^*\big]_{11},
\end{split} \\
\begin{split}
    \tilder_{ijk}(\omega_+,\omega_-) =&\ (-1)^{i+j+k-1}  \theta_{{\rm SH}i} \theta_{{\rm SH}j} \theta_{{\rm SH}k} \lambda_{{\rm N}k} \sigma_{{\rm N}j} \\
    &\, \times \sum_{l=1}^{2} \tilde \calzzeta_{il}(\Omega) \big[ \eta_{lj\perp}(\omega_+) \zzeta_{lk\perp}(\omega_-)^* \\
    &\, + \mathcal{Z}_{{\rm FN}l\parallel} \eta_{lj\perp}(\omega_+) Z_{{\rm N}l}^{-1} \tilde \zzeta_{lk\perp}(\omega_-)^* \big]_{11}.
\end{split}
\end{align*}

Being cubic in the spin-Hall angles $\theta_{{\rm SH} i}$, the spin-torque contribution is typically smaller than the Joule-heating and phonon-mediated contributions to the bilinear response. This smallness can be compensated, however, if the frequencies $\Omega$ or $\omega_{\pm}$ are equal to a resonance frequency 
\begin{equation}
    \omega_n = \Dex \left( \frac{n \pi}{d_{\rm F}} \right)^2 + \omega_0,
\label{eq:resonancefrequencies}
\end{equation}
of F, where the transverse linear-response coefficients $\zzeta_{{ij\perp}}$ and $\eta_{ij\perp}$ are resonantly enhanced \cite{Franke2025-lin}. This resonant enhancement occurs for the coefficients $\tildew_{ijk}(\omega_+,\omega_-)$ for $\Omega = \omega_n$ or $\omega_+ = \omega_n$ and for  $\tilder_{ijk}(\omega_+,\omega_-)$ if $\omega_+ = \omega_n$ or $\omega_- = \omega_n$. For these frequencies, the spin-torque contribution is the dominant source of nonlinear response, as we discuss in detail in the next Section.

{\renewcommand{\arraystretch}{1.3}
\begin{table}
\begin{ruledtabular}
    \centering
    \begin{tabular}{llr}
    \textrm{Quantity} & \textrm{Value} & \textrm{Ref.} \\
    \colrule
    $T$ & $ 300 \, \mathrm{K}$ & - \\
    $g_{\uparrow\downarrow} $ & $(6 + 0.3i) \times 10^{13} \, \Omega^{-1} \, \mathrm{m}^{-2}$ & \onlinecite{Qiu2013-rv, Althammer2013-zm, Hahn2013-rw} \\
    \hline
    \multicolumn{3}{c}{YIG} \\
    \hline
    $d_{\rm F}$ & $6 \times 10^{-8} \, \mathrm{m}$ & - \\
    $\omega_0 / 2 \pi$ & $ 8 \times 10^{9} \, \mathrm{s}^{-1}$ & \onlinecite{Hahn2013-rw} \\
    $D_{\rm ex}$ & $8 \times 10^{-6} \, \mathrm{m}^2 \, s^{-1}$ & \onlinecite{Weiler2013-ny} \\
    $s$ & $ 5.28 \times 10^{27} \, \mathrm{m}^{-3}$ & \onlinecite{Cherepanov1993-sr} \\
    \hline
    \multicolumn{3}{c}{Pt} \\
    \hline
    $d_{\rm N}$ & $ 4 \times 10^{-9} \, \mathrm{m}$ & - \\
    $\theta_{\rm SH}$ & $ 0.1 $ & \onlinecite{Althammer2013-zm} \\
    $\lambda_{\rm N}$ & $ 2 \times 10^{-9} \, \mathrm{m}$ & \onlinecite{Weiler2013-ny} \\
    $\sigma_{\rm N}$ & $ 9 \times 10^{6} \, \Omega^{-1} \, \mathrm{m}^{-1}$ & \onlinecite{Corti1984-xd} \\
    $C_{\rm p}$ & $ 2.73 \times 10^{6} \, \mathrm{J} \, \mathrm{K}^{-1} \, \mathrm{m}^{-3}$ & \onlinecite{Lin2008-tz, Rumble2022-ev} \\
    $C_{\rm e}$ & $ 0.13 \times 10^{6} \, \mathrm{J} \, \mathrm{K}^{-1} \, \mathrm{m}^{-3}$ & \onlinecite{Lin2008-tz, Rumble2022-ev, Sullivan2023-mw} \\
    $l_{\rm ep}$ & $ 4.5 \times 10^{-9} \, \mathrm{m}$ & \onlinecite{Flipse2014-kh, Weiler2013-ny} \\
    $\alpha_{{\rm TCR}}$ & $1.2 \times 10^{-3} \, \mathrm{K}^{-1}$ & \onlinecite{Sullivan2023-mw} \\ 
    $\tau_{\rm p}$ & $1 \, \mathrm{ps}$ & - \\ \hline \hline
    $|v_{111}^{\rm dc, Jo}(0)|$ & $1.78 \times 10^{-19} \, \mathrm{m}^2 \, \mathrm{V}^{-1}$ & Eqs.~\eqref{eq:joule} and \eqref{eq:v111dc} \\
    \end{tabular}
    \caption{Material parameters for YIG and Pt. Additional parameters for YIG can be found in Tab.\ \ref{tab:derivedparameters}. The electron-phonon time in Pt, $\tau_{\rm ep} \approx 40 \, \mathrm{fs}$ in Eq.~\eqref{eq:heatcontinuity}, is calculated from $l_{\rm ep}$ and $\kappa_{\rm e}$, which is in turn calculated from $\sigma_{\rm N}$ via the Wiedemann-Franz law. The bilinear response coefficients in Figs.\ \ref{fig:x111}--\ref{fig:xdf} are normalized by $v^{\rm dc,Jo}_{111}(0)$, which is calculated from the parameters listed in the Table using Eqs.\ (\ref{eq:v111dc}) and (\ref{eq:joule}).}
    \label{tab:materialparameters}
\end{ruledtabular}
\end{table}}

\begin{figure*}[t]
     \centering
     \includegraphics[width=0.9\textwidth]{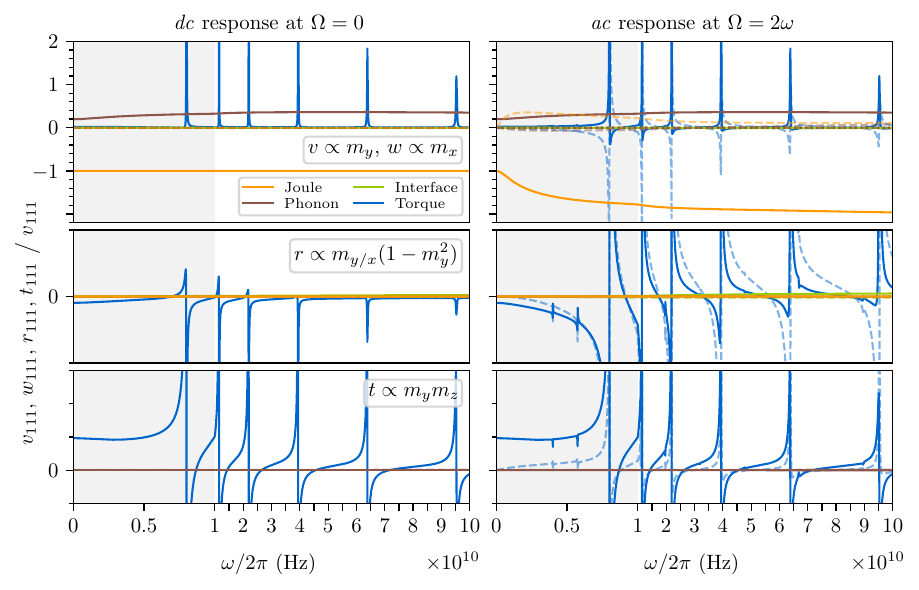}
     \caption{Real part (solid lines) and imaginary part (dashed lines) of local nonlinear response coefficients from Joule-heating (orange), phonon-mediated (brown), interfacial (green), and spin-torque (blue) contributions for a Pt$|$YIG$|$Pt trilayer, vs.\ driving frequency $\omega$, with material and device parameters taken from Tab.\ \ref{tab:materialparameters}. The left panels show the dc response at frequency $\Omega = 0$, the right panels show the ac response at $\Omega = 2 \omega$. The response coefficients are normalized by the local zero-frequency Joule-heating coefficient $v^{\rm dc,Jo}_{111}(0)$.}
     \label{fig:x111}
\end{figure*}

\section{Numerical estimates}
\label{sec:numericalestimates}

We now evaluate and compare the bilinear response coefficients $v_{ijk}(\omega_+,\omega_-)$, $w_{ijk}(\omega_+,\omega_-)$,  $r_{ijk}(\omega_+,\omega_-)$, and $t_{ijk}(\omega_+,\omega_-)$ for typical parameters for a Pt$|$YIG$|$Pt trilayer. Hereto, we consider a harmonic driving field in N1,
\begin{equation}
  E_1(t) = 2 E \cos(\omega t),\ \ E_2(t) = 0,
\end{equation}
so that the quadratic-in-$E$ current response has components at frequencies $\Omega = 0$ and $\Omega = 2 \omega$,
\begin{align}
  \bar i_{i}^{x}(t)^{(2)} =&\ 
  \sigma_{i11}^{x, {\rm dc}} E^2 + \mbox{Re}\, \sigma_{i11}^{x, {\rm ac}} E^2 e^{-2 i \omega t}
\label{eq:ix_quadratic_response}
  , \\
  \bar i_{i}^{y}(t)^{(2)} =&\ \sigma_{i11}^{y, {\rm dc}} E^2 + \mbox{Re}\, \sigma_{i11}^{y, {\rm ac}} E^2 e^{-2 i \omega t},
\label{eq:iy_quadratic_response}
\end{align}
with real response coefficient $\sigma_{i11}^{x/y, {\rm dc}} = \sigma^{x/yxx(2)}_{i11}(\omega,\omega) + \sigma^{x/yxx(2)}_{i11}(-\omega,-\omega)$ and complex response coefficients $\sigma_{i11}^{x/y, {\rm ac}} = 2 \sigma^{x/yxx(2)}_{i11}(\omega,-\omega)$. Hence, the bilinear response at frequencies $\Omega = 0$ and $\Omega = 2 \omega$ is governed by the dimensionless response coefficients
\begin{align}
\begin{split}
  \label{eq:v111dc}
  v_{i11}^{\rm dc} =&\, v_{i11}(\omega,\omega) + v_{i11}(-\omega,-\omega), \\
  v_{i11}^{\rm ac} =&\, 2 v_{i11}(\omega,-\omega),
\end{split}
\end{align}
with analogous definitions for $w_{i11}^{\rm dc/ac}$, $r_{i11}^{\rm dc/ac}$, and $t_{i11}^{\rm dc/ac}$. 

\begin{figure*}[t]
     \centering
     \includegraphics[width=0.9\textwidth]{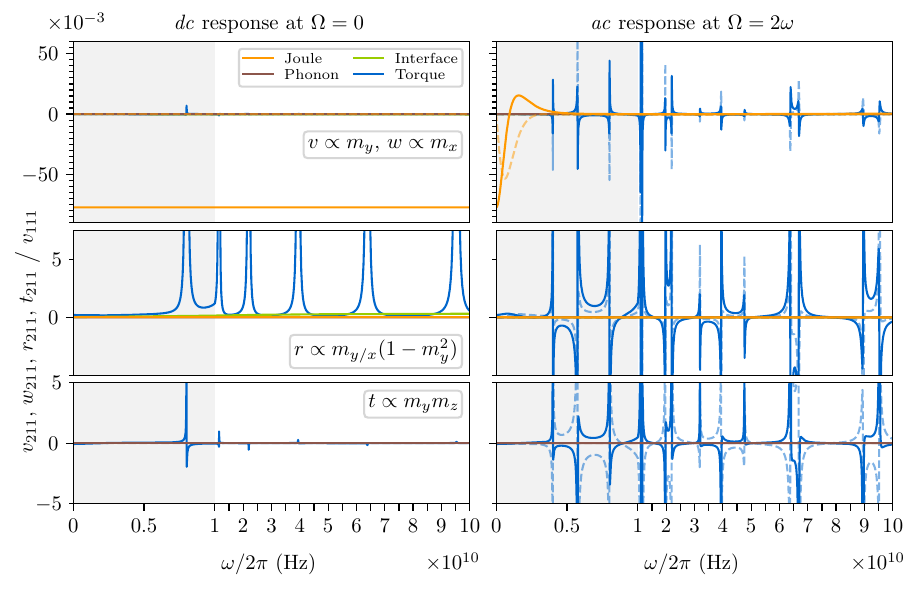}
     \caption{Same as Fig.\ \ref{fig:x111}, but for the nonlocal bilinear response coefficients, which describe the quadratic-in-$E$ charge current in N2 for an applied electric field in N1. (The response coefficients are again normalized by the local coefficient $v^{\rm dc,Jo}_{111}(0)$.)}
     \label{fig:x211}
\end{figure*}

\begin{figure}[t]
     \centering
     \includegraphics[width=0.5\textwidth]{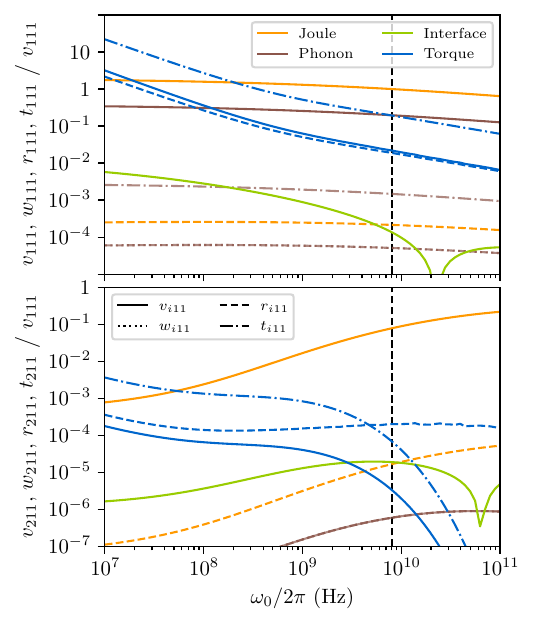}
     \caption{Local (top) and nonlocal (bottom) bilinear response coefficients in the limit of low driving frequency $\omega \to 0$ as a function of the ferromagnetic resonance frequency $\omega_0$. Device and parameter values other than $\omega_0$ are taken from Tab.\ \ref{tab:materialparameters}; the value for $\omega_0$ used in other plots is denoted as a black line.
  \label{fig:xb}}
\end{figure}

\begin{figure}[t]
     \centering
     \includegraphics[width=0.5\textwidth]{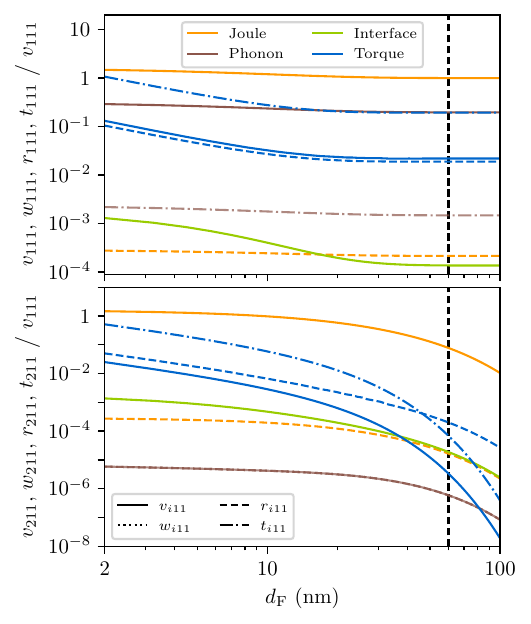}
     \caption{Local (top) and nonlocal (bottom) bilinear response coefficients in the limit of low driving frequency $\omega \to 0$ as a function of the thickness $d_{\rm F}$ of the ferromagnetic layer. The value for $d_{\rm F}$ used in other plots is denoted as a black line.
  \label{fig:xdf}}
\end{figure}

In Figs.\ \ref{fig:x111}--\ref{fig:xdf} we compare the four bilinear response coefficients $v_{i11}^{\rm dc/ac}$, $w_{i11}^{\rm dc/ac}$, $r_{i11}^{\rm dc/ac}$, and $t_{i11}^{\rm dc/ac}$ that describe local ($i=1$) and nonlocal ($i=2$) response with typical parameters for a Pt$|$YIG$|$Pt trilayer, for all four bilinear response mechanisms considered in Sec.\ \ref{sec:nonlinearresponse}. We make use of the expressions for the linear response coefficients $\zzeta_{ij\perp}$ and $\calzzeta_{ij}$, which enter into the expressions of Sec.\ \ref{sec:nonlinearresponse}, given in Ref.\ \onlinecite{Franke2025-lin} and App.\ \ref{sec:impedances}. The material and device parameters used are summarized in Tab.~\ref{tab:materialparameters}. For the Joule-heating and phonon-mediated USMR, we also include contributions not included in Sec.\ \ref{sec:nonlinearresponse}, such as the Joule-heating from dissipative spin currents in N. These contributions are calculated in App.\ \ref{app:bilinear}. The response coefficients shown in Figs.\ \ref{fig:x111}--\ref{fig:xdf} are normalized to the local response coefficient for Joule heating $v_{111}^{\rm dc}$, which is the largest response coefficient for the Pt$|$YIG$|$Pt trilayer in the {\it dc} limit.
Figures \ref{fig:x111} and \ref{fig:x211} show the frequency dependence of the coefficients for local and nonlocal bilinear response, respectively. The dependence on ferromagnetic resonance frequency $\omega_0$, which depends on the applied magnetic field, and the thickness $d_{\rm F}$ of the F layer is illustrated in Figs.\ \ref{fig:xb} and \ref{fig:xdf} for the limit of low driving frequency $\omega$. (The parameters $\omega_0$ and $d_{\rm F}$ enter our calculations through the linear-response coefficients $\calzzeta_{ij}$ and $\zzeta_{ij\perp}$, see Ref.\ \onlinecite{Franke2025-lin} and App.\ \ref{sec:impedances}. The ferromagnetic resonance frequency $\omega_0$ also enters through the magnon density of states, which determines the transport of spin and heat through the F$|$N interfaces, see Eq.\ (\ref{eq:num}).)

{\em Low-frequency limit.---} The contributions from Joule heating or phonon-mediated USMR, which are linear in the spin-Hall angle $\theta_{\rm SH}$, dominate the local low-frequency bilinear response, except in the limit of small ferromagnetic resonance frequency $\omega_0$ and/or F-thickness $d_{\rm F}$, where the spin-torque contribution dominates. The Joule-heating and phonon-mediated USMR can easily be distinguished by comparison of the $xxx$ versus $yxx$ response: while the Joule-heating contribution is of similar magnitude in both cases, the phonon-mediated USMR has a significant $xxx$ response only ~\cite{Sullivan2023-mw}. The shift from Joule-heating dominated to spin-torque dominated bilinear response comes with a shift of the dominant magnetization dependence from $\sigma^{x/yxx(2)} \propto m_{y/x}$ to a pure Hall response $\sigma^{yxx(2)} \propto m_y m_z$ (corresponding to a shift from response coefficients $v$ and $w$ to $t$, see Eqs.\ (\ref{eq:sigmaxxx}) and (\ref{eq:sigmayxx})). The interfacial USMR contribution, which is cubic in the spin-Hall angle \cite{Sterk2019-mt}, is smaller by at least three orders of magnitude for all parameter values considered. 

The nonlocal response in the low-frequency limit is dominated by the Joule-heating contribution for almot the full parameter range considered, except for the regime of very small $\omega_0$ and $d_{\rm F}$, where the spin-torque USMR contribution takes over. For the non-local response, the phonon-mediated and interfacial USMR are smaller than the leading contribution by at least three orders of magnitude.

The increase of the spin-torque contributions upon decreasing $\omega_0$ shown in Fig.\ \ref{fig:xb} reflects the increased susceptibility of the magnetization in this limit. On the other hand, the local Joule-heating and interfacial USMR contributions, which rely on thermal magnons, depend only weakly on $\omega_0$ for small anisotropies but decrease in the limit of very strong magnetic fields \cite{Sterk2019-mt}. This is different for the nonlocal response, because an increase in $\omega_0$ has a strong effect on the magnon relaxation lengths and, hence, leads to a strong increase of those contributions to the bilinear response that involve the transport of incoherent magnons across the ferromagnetic layer. 

The $1/d_{\rm F}$-scaling of the spin-torque contribution at low driving frequency shown in Fig.\ \ref{fig:xdf} sets in for $|k(0)| d_{\rm F} \lesssim 1$, where $k(0) = i \sqrt{\omega_0/\Dex}$. Our findings support the finding of \citet{Sterk2019-mt} that the local interfacial USMR is independent of $d_{\rm F}$ in most parameter regimes. The monotonic field dependence of the interfacial USMR is specific to ferromagnetic insulators, whereas \citet{Cheng2023-zg} show that the antiferromagnetic USMR inherits the nonmonotonic field dependence of the antiferromagnetic magnon numbers.

{\em Frequency dependence.---} The Joule-heating, phonon-mediated, and interfacial USMR contributions involve longitudinal spin transport only, which is mediated by incoherent (thermal) magnons. Correspondingly, these are smooth functions of the driving frequency $\omega$. The spin-torque contribution to the bilinear response also involves coherent magnetization modes. It shows sharp features as a function of the driving frequency $\omega$ if $\omega$ equals a resonance frequency $\omega_n$, see Eq.\ (\ref{eq:resonancefrequencies}), or if $\omega = \omega_n/2$, whereby for the local response the resonant features at $\omega = \omega_n/2$ are weaker than those at $\omega = \omega_n$, see Fig.\ \ref{fig:x111}.
Away from the resonance frequencies, the  $xxx$ bilinear response for all $\vm_{\rm eq}$ and the $yxx$ response for $m_z = 0$ are dominated by the Joule-heating and phonon-mediated USMR contributions. 
In the vicinity of the resonance frequencies, the bilinear response is dominated by the spin-torque contribution. It has sharp resonant features, whereby the coefficients $v_{ijk}^{\rm to} = w_{ijk}^{\rm to}$, $r_{ijk}^{\rm to}$, and $t_{ijk}^{\rm to}$ describing contributions with different characteristic magnetization dependences all have a comparable magnitude at $\omega \approx \omega_n$.
Additional resonances appear at $\omega_n/2$ in the {\it ac} response.

In App.\ \ref{app:estimates} we give analytical order-of-magnitude estimates for the dimensionless bilinear response coefficients.

\section{Conclusion}
\label{sec:conclusion}

In this work, we presented a theory of quadratic-in-applied-field response of bilayers and trilayers of magnetic insulators and normal metals, using the linear-response theory of our companion article \cite{Franke2025-lin} as a starting point. Our theory takes into account four sources such unidirectional spin-Hall magnetoresistance (USMR) that were previously considered in the literature --- Joule heating from charge and spin currents \cite{Tulapurkar2011-lm, Taniguchi2016-ei}, phonon-mediated unidirectional magnetoresistance \cite{Sullivan2023-mw}, spin-orbit torque \cite{Avci2018-uv, Liu2021-ny}, and a magnonic interfacial contribution \cite{Sterk2019-mt, Liu2021-ny, Avci2018-uv} --- and allows a quantitative comparison of these four mechanisms over a broad range of frequencies up to the THz regime. We identify the characteristic magnetization-direction dependence of each of the four mechanisms and show that the dominant bilinear effect changes with the driving frequency of the applied electric field (see Figs.~\ref{fig:x111} and \ref{fig:x211}). In addition, the individual contribution of each of the effects to the total bilinear response changes with external magnetic field and geometry and can be vastly different in a local or nonlocal measurement (see Figs.~\ref{fig:xb} and \ref{fig:xdf}). These results offer key insights for experimental differentiation of nonlinear effects and underscore their potential for advancing nonlinear spintronic applications.

We briefly highlight some findings.
Many experiments observe Joule heating effects as the dominant second-harmonic response to an {\it ac} driving field $E$ \cite{Althammer2021-al, Cornelissen2018-dl}. 
There are, however, parameter regimes where Joule heating is not the dominant contribution to the quadratic-in-$E$ response. Strong electron-phonon coupling in the normal metal N combined with a small phonon heat capacity leads to a large temperature change mediated by the spin-Peltier effect \cite{Sullivan2023-mw}. Since the electrical conductivity of the normal-metal layers depends on phonon temperature, this gives rise to a phonon-mediated unidirectional magnetoresistance, which can be of comparable magnitude to the unidirectional response from Joule heating. Other parameter regimes in which Joule heating may not be the dominant contribution to the quadratic-in-$E$ response include small magnetic anisotropies, small thicknesses of the magnetic insulator F, and a driving frequency close to resonant frequencies of coherent magnons. In these instances, the spin-torque diode effect is expected to dominate \cite{Avci2017-kj, Liu2021-ny}. A subordinate role throughout the entire parameter range is played by the interfacial magnonic USMR, which we find to be at least two orders of magnitude smaller than the Joule heating or spin-torque contributions in the {\it dc} limit. The USMR effect is significantly larger in magnetic metals \cite{Avci2015-pk}, where the largest contribution is from the electronic spin accumulation not present in insulators. In addition, the latter have a smaller thermal conductivity, which results in inhomogeneous heating in F and N and ultimately to Joule heating as the dominant second-order-in-$E$ response \cite{Sterk2019-mt}.

Since all four bilinear contributions change sign under reversal of the magnetization direction, they constitute a unidirectional magnetoresistance (UMR) in F$|$N multilayers. Unidirectional effects are also present in heterostructures with antiferromagnetic-insulator \cite{Shim2022-zn, Zheng2023-kf, Wang2024-cx, Cheng2023-zg}, topological-insulator \cite{Lv2018-rq}, or magnetic-metal \cite{Avci2015-pk, Avci2018-uv} layers and their origins differ between those systems. Recent years have shown a particular interest in antiferromagnetic UMR, which is driven by both magnon dynamics --- altered by a field-induced spin canting --- and an interfacial Rashba splitting \cite{Shim2022-zn, Zheng2023-kf}. We note that antiferromagnetic UMR effects show a different experimental signature compared to the ferromagnetic effects discussed here, especially with regard to its dependence on external magnetic fields.

From an applied perspective, bilinear mechanisms offer additional functionality in spintronic devices. Their unidirectional nature enables the detection of magnetization switching in a two-terminal setup \cite{Avci2015-pk} or multi-state memory devices \cite{Avci2017-xh}. Controlling unidirectional response on ultrafast time scales may be an important step towards efficient and fast spintronic information processing.

\section*{Acknowledgments}
We thank U. Gems, D. A. Reiss, and T. Kampfrath for stimulating discussions. This work was funded by the Deutsche Forschungsgemeinschaft (DFG, German Research Foundation) through the Collaborative Research Center SFB TRR 227 ``Ultrafast spin dynamics'' (Project-ID 328545488, project B03).

\begin{appendix}
\renewcommand{\thefigure}{A\arabic{figure}}
\setcounter{figure}{0}
\renewcommand{\thetable}{A\arabic{table}}
\setcounter{table}{0}

\section{Linear response coefficients}
\label{sec:impedances}
In this Appendix, we summarize the linear response of an N$|$F$|$N trilayer to an applied field, using the formalism of Ref.~\cite{Franke2025-lin}. In the main text, the linear response is captured by the dimensionless response coefficients $\zzeta_{ij\perp}(\omega)$ and $\calzzeta_{ij}(\omega)$, see Eq.~\eqref{eq:linresponse}.
These coefficients describe the response of the entire N$|$F$|$N trilayer, including the central F part, which is not considered explicitly in the main text. Referring to Ref.\ \onlinecite{Franke2025-lin} for a complete description, we here summarize the essentials. 

In Ref.\ \onlinecite{Franke2025-lin}, coherent magnon transport through F, which contributes to the transverse component $i_{{\rm s}\perp}(z,t)$ of the spin current, is described for magnons with a dispersion 
\begin{equation}
  k(\omega) = \sqrt{\frac{\omega(1+i\alpha)-\omega_0}{\Dex}},
\end{equation}
where $\omega_0$ is the ferromagnetic resonance frequency, $\Dex$ the spin stiffness, and $\alpha$ the phenomenological Gilbert damping coefficient. For the response coefficient $\zzeta_{ij\perp}(\omega)$, Ref.\ \onlinecite{Franke2025-lin} then finds
\begin{align}
  \zzeta_{{ji\perp}}(\omega) =&\ \delta_{ji} - (-1)^{i+j} \frac{Z_{{\rm N}j}}{Z_{ji\perp}(\omega)},
  \label{eq:zetaperp}
\end{align}
where
\begin{widetext}
\begin{align}
\begin{split}
  Z_{11\perp}(\omega) =&\ Z_{{\rm N}1} + g_{\uparrow\downarrow 1}^{-1}
  + \left[\cos(k(\omega) d_{\rm F}) (Z_{{\rm N}2} + g_{\uparrow\downarrow 2}^{-1})
  -i \sin(k(\omega) d_{\rm F}) Z_{{\rm F}\perp}^{\infty}(\omega) \right]
  \\ &\, \times 
  \left[ \cos(k(\omega) d_{\rm F}) Z_{{\rm F}\perp}^{\infty}(\omega) - i \sin(k(\omega) d_{\rm F})
  (Z_{{\rm N}2} + g_{\uparrow\downarrow 2}^{-1}) \right]^{-1}  Z_{{\rm F}\perp}^{\infty}(\omega),
  \label{eq:Z11perp}
\end{split} \\
\begin{split}
  Z_{21\perp}(\omega) =&\
  \left(Z_{{\rm N}1} + g_{\uparrow\downarrow 1}^{-1} \right) \left(Z_{{\rm F}\perp}^{\infty}(\omega) \right)^{-1}
  \left[ \cos(k(\omega) d_{\rm F}) Z_{{\rm F}\perp}^{\infty}
    - i \sin(k(\omega) d_{\rm F}) (Z_{{\rm N}2} + g_{\uparrow\downarrow 2}^{-1}) \right] \\
    &\, + \cos(k(\omega) d_{\rm F}) (Z_{{\rm N}2} + g_{\uparrow\downarrow 2}^{-1}) - i \sin(k(\omega) d_{\rm F}) Z_{{\rm F}\perp}^{\infty}(\omega)
  \label{eq:Z21perp}
\end{split} 
\end{align}
\end{widetext}
with $Z_{{\rm N}i}$ defined in Eq.\ (\ref{eq:zn}) and 
\begin{equation}
  Z_{{\rm F}\perp}^{\infty}(\omega) = \frac{\hbar \omega}{2 e^2 \Dex k(\omega) s}.
\label{eq:zfperpinf}
\end{equation}
The effective impedances $Z_{12\perp}(\omega) = Z_{21\perp}(\omega)$ and $Z_{22\perp}(\omega)$ is obtained from Eqs.~(\ref{eq:Z11perp}) and (\ref{eq:Z21perp}) by interchanging the indices $1 \leftrightarrow 2$. 

Incoherent magnons, which contribute to the longitudinal component $i_{{\rm s}\parallel}(z,t)$ of the spin current and to the heat current, are described with the help of $2 \times 2$ generalized conductivity and heat capacity matrices $\Sigma_{\rm m}(\omega)$ and ${\cal C}_{\rm m}$ \cite{Franke2025-lin},
\begin{align}
  \Sigma_{\rm m}(\omega) =&\, \frac{1}{1 - i \omega \tau_{\rm m}}
    \begin{pmatrix}
        \sigma_{\rm m} & e L_{\rm m}/k_{\rm B} T \\
        e L_{\rm m}/k_{\rm B} T & 2 e^2 \kappa_{\rm m} / k_{\rm B}^2 T
    \end{pmatrix}
    \label{eq:capitalsigma} \\
  {\cal C}_{\rm m} =&\, \begin{pmatrix} c_{{\rm m,s}\mu} & c_{{\rm m,s}T} \\
  c_{{\rm m,Q}\mu} & c_{{\rm m,Q}T} \end{pmatrix}
\end{align}
where $\tau_{\rm m}$ is the magnon momentum relaxation time, $\sigma_{\rm m}$ the magnon conductivity, $L_{\rm m}$ the spin-Seebeck coefficient, $\kappa_{\rm m}$ the magnon thermal conductivity, $c_{{\rm m,s}\mu} = (2 e/\hbar) \partial \rho_{\rm ms}/\partial u_{\rm m}$, $c_{{\rm m,s}T} = c_{{\rm m,Q}\mu} = (2 e/\hbar) \partial \rho_{\rm ms}/\partial \Delta u_{\rm mQ}$, and $c_{{\rm m,Q}T} = (2 e/k_{\rm B} T) \partial \rho_{\rm ms}/\partial \Delta u_{\rm mQ}$, with $\rho_{\rm ms}$ and $\rho_{\rm mQ}$ the magnonic angular-momentum and energy density, respectively. Together, these matrices determine the relaxation lengths of coupled magnon spin and heat transport via the $2 \times 2$ matrix \cite{Franke2025-lin}
\begin{align}
  \label{eq:Lambda}
  \Lambda^2(\omega) =&\, 
  \left[  \frac{1 - i \omega \tau_{{\rm mp,rel}}}{\tau_{{\rm mp,rel}}}
  {\cal C}_{\rm m} 
  + (1-\gamma) 
  \begin{pmatrix} 
  \frac{c_{{\rm m,s}\mu}}{\tau_{\rm m,rel}} & 0 \\ 0 &
  \frac{c_{{\rm m,Q}T}}{\tau_{\rm mp,ex}} \end{pmatrix} \right]^{-1}
  \nonumber \\ &\, \mbox{} \times \Sigma(\omega),
\end{align}
with $\tau_{\rm mp,rel}$, $\tau_{\rm m,rel}$, and $\tau_{\rm mp,ex}$ the relaxation times for spin-non-conserving magnon-phonon scattering, spin-non-conserving magnon-magnon scattering, and spin-conserving magnon-phonon scattering, respectively, and
\begin{equation}
  \gamma = \frac{c_{{\rm m,s}T} c_{{\rm m,Q}\mu}}{c_{{\rm m,s}\mu} c_{{\rm m,Q}T}}.
\end{equation}
In Ref.\ \onlinecite{Franke2025-lin}, it is then shown that
\begin{align}
  \label{eq:zetapara}
  \calzzeta_{{ji}}(\omega) =&\,
  \mathbb{I}_2 \delta_{ji} - (-1)^{i+j} {\cal Z}_{{\rm N}j}(\omega) \mathcal{Z}_{ji\parallel}^{-1} (\omega),
\end{align}
with
\begin{widetext}
\begin{align}
\begin{split}
  \mathcal{Z}_{11\parallel}(\omega) =&\
  \mathcal{Z}_{{\rm N}1} + \mathcal{Z}_{{\rm FN}1\parallel}
  \, + \left[ \sinh(\Lambda(\omega)^{-1} d_{\rm F}) 
  \Lambda(\omega) \Sigma_{\rm m}(\omega)^{-1}
    + \cosh(\Lambda(\omega)^{-1} d_{\rm F})
    (\mathcal{Z}_{{\rm N}2} + \mathcal{Z}_{{\rm FN}2\parallel})
  \right]  \\ &\, \times
   \left[\cosh(\Lambda(\omega)^{-1} d_{\rm F}) 
  \Lambda(\omega) \Sigma_{\rm m}(\omega)^{-1}
     + \sinh(\Lambda(\omega)^{-1} d_{\rm F})
     (\mathcal{Z}_{{\rm N}2} + \mathcal{Z}_{{\rm FN}2\parallel})
   \right]^{-1}
  \Lambda(\omega) \Sigma_{\rm m}(\omega)^{-1}
\label{eq:calZ11para}
\end{split} \\
\begin{split}
  \mathcal{Z}_{21\parallel}(\omega) =&\
  (\mathcal{Z}_{{\rm N}1} + \mathcal{Z}_{{\rm FN}1\parallel})
  \Sigma_{\rm m}(\omega) \Lambda(\omega)^{-1}
  \left[\cosh(\Lambda(\omega)^{-1} d_{\rm F}) 
  \Lambda(\omega) \Sigma_{\rm m}(\omega)^{-1}
    + \sinh(\Lambda(\omega)^{-1} d_{\rm F})
    (\mathcal{Z}_{{\rm N}2} + \mathcal{Z}_{{\rm FN}2\parallel})
    \right] 
   \\ &\,
   + \sinh(\Lambda(\omega)^{-1} d_{\rm F}) 
  \Lambda(\omega) \Sigma_{\rm m}(\omega)^{-1}
   + \cosh(\Lambda(\omega)^{-1} d_{\rm F}) 
  (\mathcal{Z}_{{\rm N}2} + \mathcal{Z}_{{\rm FN}2\parallel}).
   \label{eq:calZ21para}
\end{split}
\end{align}
\end{widetext}
The impedances $Z_{{\rm N}i}$, $\mathcal{Z}_{{\rm N}i}$, and $\mathcal{Z}_{{\rm FN}i\parallel}$ are defined in Eqs.~\eqref{eq:zn}, \eqref{eq:znmat}, and \eqref{eq:calzfnpara}, respectively. The impedance matrices $\mathcal{Z}_{12\parallel}(\omega)$ and $\mathcal{Z}_{22\parallel}(\omega)$ are obtained from Eqs.~(\ref{eq:calZ11para}) and (\ref{eq:calZ21para}) by interchange of the indices. 

Typical values for the additional parameters appearing in this appendix for magnons in YIG are summarized in Tab.\ \ref{tab:derivedparameters}.

{\renewcommand{\arraystretch}{1.3}
\begin{table}
\begin{ruledtabular}
    \centering
    \begin{tabular}{lllr}
    \multicolumn{2}{l}{\textrm{Quantity}} & \textrm{Value} & \textrm{source} \\
    \colrule
    $\sigma_{\rm m}$ & & $4.1 \times 10^5 \, \mathrm{S} \, \mathrm{m}^{-1}$
  & \textrm{Ref.\ \onlinecite{Cornelissen2015-fh}} \\
    $L_{\rm m}$ & & $1.4 \times 10^4 \, \mathrm{A} \, \mathrm{m}^{-1}$ & \onlinecite{Franke2025-lin}\\
    $\kappa_{\rm m}$ & & $1.8 \, \mathrm{W} \, \mathrm{K}^{-1} \, \mathrm{m}^{-1}$ & \onlinecite{Franke2025-lin} \\
    $\alpha$ & & $2 \times 10^{-4}$ & \onlinecite{Cornelissen2015-fh} \\
    $\tau_{\rm m}$ & & $ 0.11 \, \mathrm{ps}$ & - \\
    $\tau_{\rm m,ex}$ & & $ 2.7 \, \mathrm{ps}$ & \onlinecite{Shi2021-kc} \\
    $\tau_{\rm m,rel}, \ \tau_{\rm mp,rel}$ & & $ 2 \hbar / (\alpha k_{\rm B} T) \approx 255 \, \mathrm{ps}$ & - \\
    \hline
    $l_{\mu}(0)$ & & $20 \, \mathrm{nm}$ & \onlinecite{Franke2025-lin} \\
    $l_{\rm T}(0)$ & & $7 \, \mathrm{nm}$ & \onlinecite{Franke2025-lin} \\
    \hline
    \multirow{4}{*}{$\mathcal{C}_{\rm m}$}
    & $c_{{\rm m,s}\mu}$ & $150 \, \mathrm{ns} \, \Omega^{-1} \, \mu\mathrm{m}^{-3}$ & \onlinecite{Franke2025-lin}\\
    & $c_{{\rm m,sT}}$ & $11 \, \mathrm{ns} \, \Omega^{-1} \, \mu\mathrm{m}^{-3}$ & \onlinecite{Franke2025-lin}\\
    & $c_{{\rm m,Q}\mu}$ & $11 \, \mathrm{ns} \, \Omega^{-1} \, \mu\mathrm{m}^{-3}$ & \onlinecite{Franke2025-lin}\\
    & $c_{{\rm m,QT}}$ & $15 \, \mathrm{ns} \, \Omega^{-1} \, \mu\mathrm{m}^{-3}$ & \onlinecite{Franke2025-lin}\\
    \hline
    \end{tabular}
    \caption{Additional parameters governing magnon transport in YIG. The relaxation lengths $l_{T}(\Omega) < l_{\mu}(\Omega)$ are the eigenvalues of $\Lambda(\Omega)$, see Eq.\ (\ref{eq:Lambda}). The impurity scattering time $\tau_{\rm m}$, the spin-Seebeck coefficient $L_{\rm m}$, and the magnon thermal conductivity $\kappa_{\rm m}$ are obtained from the magnon conductivity $\sigma_{\rm m}$ via Drude-Boltzmann theory, see Ref.\ \onlinecite{Franke2025-lin}. The time scale $\tau_{\rm m,ex}$ is mainly associated with ``four-magnon scattering'' and we estimate the combined $\tau_{\rm m,rel}$ and $\tau_{\rm mp,rel}$ from the phenomenological Gilbert damping constant, since both processes are not magnon-number conserving. At room temperature, spin-non-conserving ``three-magnon scattering'' can be neglected in comparison to the spin-conserving ``four-magnon scattering'' \cite{Shi2021-kc}. 
}
    \label{tab:derivedparameters}
\end{ruledtabular}
\end{table}}

\section{Bilinear response}
\label{app:bilinear}
In this appendix, we present additional details for the calculation of the nonlinear response in Sec.~\ref{sec:nonlinearresponse}.

\subsection*{Joule heating contribution}
The Joule heating rate is \cite{Tulapurkar2011-lm, Taniguchi2016-ei}
\begin{align}
\begin{split}
  s(\vr,t) =&\,
  - \frac{1}{e} \vnabla \cdot (\vi(\vr,t) \cdot \varphi_{\rm c}(\vr,t))
  \\ &\ \mbox{}
  - \frac{1}{\hbar} \vnabla \cdot (\mathbf{j}_{\rm s}(\vr,t) \cdot \vmus(\vr,t)),
\end{split}
\label{eq:sJoule}
\end{align}
where $\varphi_{\rm c}(\vr,t)$ is the electrochemical potential, $\mathbf{j}_{\rm s}(\vr,t) = (\hbar/2 e) \vis(\vr,t)$ the spin current tensor, and $\vmus(\vr,t) = e \vu_{\rm s}(\vr,t)$ the spin accumulation. The charge current density $i^z(\vr,t) = 0$, whereas $i^x(\vr,t)$ and $i^y(\vr,t)$ depend on $z$ only. The electrochemical potential $\varphi_{\rm c}(\vr,t)$ does not depend on $y$. Hence, the first term in Eq.~(\ref{eq:sJoule}) simplifies to $-(1/e) (i^x (z,t) \partial \varphi_{\rm c}(\vr,t)/\partial x = i^x(z,t) E_i(t)$. For the second term, we note that both $\vis(\vr,t)$ and $\vus(\vr,t)$ depend on $z$ only, so that Eq.~(\ref{eq:jouleheatingrate}) of the main text follows.

Using Eqs.~(\ref{eq:ix})--(\ref{eq:spincontinuity}), the charge and spin current densities in N1 can be expressed in terms of the applied field $E_1(t)$ and the spin accumulation $\vu_{{\rm s}1}(t)$ at the F$|$N1 interface at $z=0$,
\begin{align}
  \vu_{{\rm s}}(z,t) =&\ \vu_{{\rm s}1}(t) e^{-z/\lambda_{{\rm N}1}}, \\
  i^x(z,t) =&\ \sigma_{{\rm N}1} E_1(t) + \frac{\theta_{{\rm SH}1} \sigma_{{\rm N}1}}{2 \lambda_{{\rm N}1}} e^{-z/\lambda_{{\rm N}1}} \vu_{{\rm s}1}(t) \cdot \ve_y, \nonumber \\ \nonumber
  \vi_{{\rm s}}^z(z,t) =&\ - \theta_{{\rm SH}1} \sigma_{{\rm N}1} E_1(t) \ve_y
  + \frac{\sigma_{{\rm N}1}}{2 \lambda_{{\rm N}1}} e^{-z/\lambda_{{\rm N}1}} \vu_{{\rm s}1}(t),
\end{align}
with analogous equations for the charge and spin current densities in N2. From Eq.~(\ref{eq:jouleheatingrate}) we then find the local Joule heating rates
\begin{align}
  \label{eq:ssol_app}
  s_1(z,t) =&\ \sigma_{{\rm N}1} E_1(t)^2 + \frac{\sigma_{{\rm N}1}}{2
  \lambda_{{\rm N}1}^2  }
  |\vu_{{\rm s}1}(t)|^2 e^{-2 z/\lambda_{{\rm N}1}}, \\ \nonumber
  s_2(z,t) =&\ \sigma_{{\rm N}2} E_2(t)^2 + \frac{\sigma_{{\rm N}2}}{2
  \lambda_{{\rm N}2}^2  }
  |\vu_{{\rm s}2}(t)|^2 e^{-2 (d_{\rm F} - z)/\lambda_{{\rm N}2}}.
\end{align}
in N1 and N2, respectively.

In our model, the Joule heating rate appears in the electron heat continuity equation \eqref{eq:heatcontinuity}. The energy injected into the electronic system raises both electron and phonon temperatures. Due to the relatively low heat capacity of electrons compared to the phonon subsystem, $C_{{\rm e}} \ll C_{{\rm p}}$, the electron temperature increases more significantly under the same energy input. In addition to electron-phonon scattering, $\tau_{{\rm ep}}$, we introduce a coupling of phonons to a bath, {\it e.g.} a substrate, $\tau_{{\rm p}}$. This leads to a quasi-equilibrium, where energy is continuously supplied by the electric field but relaxation processes maintain a steady-state temperature difference between electrons and phonons. The temperature ratio is determined by the ratios of the heat capacities and the relaxation times, see Eq.~\eqref{eq:gi}, which follows from two coupled rate equations for electron and phonon temperatures. In the following, we take the limit $C_{{\rm e}i} \ll C_{{\rm p}i}$, so that $\Delta T_{\rm p}$ may be neglected when calculating the Joule-heating contribution. (For the phonon-mediated contribution, we consider a finite $\Delta T_{\rm p}$ given by Eq.~\eqref{eq:gi}.)

Inserting Eq.~(\ref{eq:ssol_app}) into Eq.~(\ref{eq:nonlin_deltauqe}) and performing a Fourier transform, the source terms $\delta u_{{\rm eQ}i}(\Omega)$ in the boundary condition (\ref{eq:nmheatrelation}) become
\begin{align}
  \label{eq:deltaT0result_app}
  \delta u_{{\rm eQ}i}(\Omega) =&\ 
  \frac{k_{\rm B}^2 l_{{\rm ep},i}(\Omega)^2 \sigma_{{\rm N}i}}{2 \pi e^2 \kappa_{{\rm e}i}}
  \int d\omega
  \left[ \vphantom{\frac{M}{M}}
    E_i(\omega_+) E_i(-\omega_-) \right. \nonumber \\  &\, \left. 
  + \frac{\vu_{{\rm s}i}(\omega_+) \cdot \vu_{{\rm s}i}(-\omega_-)}{4 \lambda_{{\rm N}i} l_{{\rm ep},i}(\Omega)} n_i(\Omega) \right],
\end{align}
where we abbreviated (assuming $d_{{\rm N}i} \gg \lambda_{{\rm N}i}$)
\begin{align}
  n_i(\Omega) =&\,
  \frac{4 l_{{\rm ep},i}(\Omega)^2 \coth(d_{{\rm N}i}/l_{{\rm ep},i}(\Omega)) -
    2  l_{{\rm ep},i}(\Omega) \lambda_{{\rm N}i}}{4 l_{{\rm ep},i}(\Omega)^2 - \lambda_{{\rm N}i}^2}.
\label{eq:jouleheatingfi}
\end{align}

The interfacial spin accumulations $\vu_{{\rm s}i}(\pm \omega_{\pm})$ can be expressed in terms of the electric field, see Eq.~(\ref{eq:linresponse}). For the source term $\delta u_{{\rm eQ}i}(\Omega)$ of Eq.~(\ref{eq:deltaT0result_app}), this gives
\begin{widetext}
\begin{align}
  \delta u_{{\rm eQ}i}(\Omega) =&\
  \sum_{j,k}
  \int \frac{{\rm d}\omega}{2 \pi} E_j(\omega_+)E_k(\omega_-)^*
  \sigma_{{\rm N}j} \sigma_{{\rm N}k}
  \frac{\lambda_{{\rm N}i} k_{\rm B} l_{{\rm ep},i}(\Omega)}{\sigma_{{\rm N}i} e \kappa_{{\rm e}i}}
  \left\{ \frac{l_{{\rm ep},i}(\Omega)}{\lambda_{{\rm N}i}}
    \delta_{ij} \delta_{ik}
    + (-1)^{j+k} \theta_{{\rm SH}j} \theta_{{\rm SH}k}
    n_i(\Omega)
    \right.
    \label{eq:dTi2} \\ &\, \times \left.
   \left[ |\vm_{\rm eq} \cdot \ve_y|^2
  \calzzeta_{ij}(\omega_+)_{11} \calzzeta_{ik}(\omega_-)^*_{11} + |\ve_{\perp} \cdot \ve_y|^2 [\zzeta_{ij\perp}(\omega_+)\zzeta_{ik\perp}(\omega_-)^*
  + \zzeta_{ij\perp}(-\omega_+)^*\zzeta_{ik\perp}(-\omega_-)] \right]
   \vphantom{\frac{M}{M}}
    \right\}. \nonumber
\end{align}
The first term proportional to $\delta_{ij} \delta_{ik}$ is the same as in Eq.~(\ref{eq:nonlin_deltauqeresponse}) of the main text. Combining Eqs.\ (\ref{eq:ix}), (\ref{eq:iy}), (\ref{eq:linresponse}), and (\ref{eq:dTi2}) we find that the charge conductivity associated with Joule heating can be written in the form of Eqs.~\eqref{eq:sigmaxxx} and \eqref{eq:sigmayxx}, where the dimensionless coefficients $v_{ijk}$, $w_{ijk}$, and $r_{ijk}$ read
\begin{align}
\label{eq:jouleB1}
\begin{split}
  v_{ijk}^{\rm Jo}(\omega_+,\omega_-) =&\ w_{ijk}^{\rm Jo}(\omega_+,\omega_-) \\
  = &\ (-1)^{i+j+k-1} \theta_{{\rm SH}i} \sigma_{{\rm N}j} \sigma_{{\rm N}k}
  \sum_{l=1}^{2} \calzzeta_{il}(\Omega)_{12} \frac{k_{\rm B}}{2 e}
  \frac{\lambda_{{\rm N}l} l_{{\rm ep},l}(\Omega)}{\sigma_{{\rm N}l} \kappa_{{\rm e}l}} \\
  &\, \times
  \left\{ 
  (l_{{\rm ep},l}(\Omega)/\lambda_{{\rm N}l})
  \delta_{lj} \delta_{lk}
  + \theta_{{\rm SH}j} \theta_{{\rm SH}k}
  n_l(\Omega)
  \calzzeta_{lj}(\omega_+)_{11} \calzzeta_{lk}(\omega_-)^*_{11}
  \right\},
\end{split} \\
\begin{split}
  r_{ijk}^{\rm Jo}(\omega_+,\omega_-) =&\, -
  \frac{1}{2} (-1)^{i+j+k-1}  \theta_{{\rm SH}i} \theta_{{\rm SH}j} \theta_{{\rm SH}k} \sigma_{{\rm N}j} \sigma_{{\rm N}k}
  \sum_{l=1}^{2} 
   \calzzeta_{il}(\Omega)_{12} \frac{k_{\rm B}}{2 e}
  \frac{\lambda_{{\rm N}l} l_{{\rm ep},l}(\Omega)}{\sigma_{{\rm N}l} \kappa_{{\rm e}l}}  n_l(\Omega) \\
  &\, \times
  \left\{ 2 \calzzeta_{lj}(\omega_+)_{11} \calzzeta_{lk}(\omega_-)^*_{11}
  - \zzeta_{lj\perp}(\omega_+)\zzeta_{lk\perp}(\omega_-)^*
  - \zzeta_{lj\perp}(-\omega_+)^*\zzeta_{lk\perp}(-\omega_-)) \right\}.
\end{split}
\label{eq:jouleB2}
\end{align}
The term proportional to $\delta_{lj} \delta_{lk}$ in Eq.~\eqref{eq:jouleB1} corresponds to Eq.~\eqref{eq:joule} of the main text.
\end{widetext}

\subsection*{Phonon-mediated contribution}
There are additional contributions to the bilinear response from phonon-mediated UMR to the ones derived in the main text that scale with the cube of the spin-Hall angle. One obtains these contributions by solving the bilinear part of the coupled differential Eqs.~\eqref{eq:ix}, \eqref{eq:iy}, \eqref{eq:jsz}, and \eqref{eq:spincontinuity} with time-dependent $\sigma_{{\rm N}i}(t)$ and $\tau_{{\rm sf},i}(t)$, which both change with phonon temperature according to Eq.~\eqref{eq:sigman_tcr}. Specifically, we insert Eq.~\eqref{eq:jsz} into \eqref{eq:spincontinuity} and consider only terms bilinear in $E$. We split products of time-dependent quantities, {\it e.g.} $\sigma_{\rm N}(z,t) \partial_z^2 \vu_{\rm s}(z,t)$, into terms proportional to the second order spin accumulation, $\sigma_{\rm N}^{(0)} \partial_z^2 \vu^{(2)}_{\rm s}(z,t)$, and products of linear-in-$E$ quantities, $\sigma_{\rm N}^{(1)}(z,t) \partial_z^2 \vu^{(1)}_{\rm s}(z,t)$. The former amounts to a diffusion equation for the spin accumulation,
\begin{equation}
    \left( \frac{ e^2 \nu_{{\rm N}i}}{\tau^{(0)}_{{\rm sf},i}} - i \Omega  e^2 \nu_{{\rm N}i}  - \frac{\sigma^{(0)}_{{\rm N}i}}{2} \frac{\partial^2}{\partial z^2} \right) \vu^{(2)}_{{\rm s}}(z,\Omega) = \boldsymbol{\xi}(z, \Omega),
\label{eq:usdiffusion_pmumr}
\end{equation}
while the latter constitutes a source term from the linear response theory,
\begin{align}
\begin{split}
    \boldsymbol{\xi}(z, t) = &\ \left( \frac{\sigma^{(1)}_{{\rm N}}(z,t)}{2} \frac{\partial^2}{\partial z^2}   - \frac{e^2 \nu_{{\rm N}i}}{\tau^{(1)}_{{\rm sf}}(z,t)} \right) \vu^{(1)}_{{\rm s}}(z,t) \\
    &\, + \theta_{{\rm SH}i} \frac{\partial}{\partial z} \sigma^{(1)}_{{\rm N}}(z,t) E_i(t) \ve_y.
\end{split}
\end{align}

The solution of Eq.~\eqref{eq:usdiffusion_pmumr} has the form of Eq.~\eqref{eq:nmspinrelation} with source voltage ({\it c.f.} Eq.~\eqref{eq:nonlin_deltauqe})
\begin{equation}
    \delta \vu_{{\rm s}1}(\Omega) = Z_{{\rm N}1} \int_0^{d_{{\rm N}1}} \mathrm{d}z' \frac{\boldsymbol{\xi}(z', \Omega) \cosh \frac{d_{{\rm N}1}-z'}{\lambda_{{\rm N}1}}}{\sinh \frac{d_{{\rm N}1}}{\lambda_{{\rm N}1}}},
\label{eq:phonon_source_integral}
\end{equation}
which gives the bilinear charge current from Eqs.~\eqref{eq:ix}, \eqref{eq:iy}, and \eqref{eq:linresponse}. We identify the dimensionless coefficients
\begin{align}
    v_{ijk}^{\rm ph}(\omega_+,\omega_-) =&\ \nu_{ijk}(\omega_+,\omega_-) + \zeta_{ijk}(\omega_+,\omega_-), \nonumber \\
    w_{ijk}^{\rm ph}(\omega_+,\omega_-) =&\, - r_{ijk}^{\rm ph}(\omega_+,\omega_-) \label{eq:phonon_appendix} \\
    = \zeta_{ijk}(\omega_+&,\omega_-) + \eta_{ijk}(\omega_+,\omega_-) + \eta_{ijk}(-\omega_+,-\omega_-)^*, \nonumber  \\
    t_{ijk}^{\rm ph}(\omega_+,\omega_-) =&\, - i \eta_{ijk}(\omega_+,\omega_-) - i \eta_{ijk}(-\omega_+,-\omega_-)^*, \nonumber
\end{align}
where we defined
\begin{widetext}
\begin{align}
\begin{split}
    \nu_{ijk}(\omega_+,\omega_-) =&\, - \delta_{ik} (-1)^{j-1} \theta_{{\rm SH}j} \calzzeta_{ij}(\omega_+)_{21} \lambda_{{\rm N}j} \frac{2 e}{k_{\rm B} } \alpha_{{\rm T},i} g_i(\omega_+) l_{{\rm ep},i}(\omega_+) \tanh{\frac{d_{{\rm N}i}}{l_{{\rm ep},i}(\omega_+)}} , 
    \\
    \eta_{ijk}(\omega_+,\omega_-) =&\ \frac{1}{2} 
    \theta_{{\rm SH}i} \theta_{{\rm SH}j} \theta_{{\rm SH}k} \lambda_{{\rm N}j} \lambda_{{\rm N}k} \frac{2 e}{k_{\rm B}} \sum_{l=1}^{2}
    \alpha_{{\rm T},l} g_l(\omega_+) 
    \zzeta_{il\perp}(\Omega) \big\{ (-1)^{i+j} \delta_{lk} n_k'(\omega_+) \calzzeta_{kj}(\omega_+)_{21} \\
    &\, + (-1)^{i+j+k-1} n_l''(\omega_+) \frac{2 l_{{\rm ep},l}(\omega_+)}{\lambda_{{\rm N}l}} \calzzeta_{lj}(\omega_+)_{21} \zzeta_{lk\perp}(-\omega_-) 
     \big\},
    \\
    \zeta_{ijk}(\omega_+,\omega_-) =&\, - \theta_{{\rm SH}i} \theta_{{\rm SH}j} \theta_{{\rm SH}k} \lambda_{{\rm N}j} \lambda_{{\rm N}k} \frac{2 e}{k_{\rm B}} \sum_{l=1}^{2} 
    \alpha_{{\rm T},l} g_l(\omega_+)
    \calzzeta_{il}(\Omega)_{11} \big\{ (-1)^{i+j} \delta_{lk}  n_k'(\omega_+) \calzzeta_{kj}(\omega_+)_{21} \\
    &\, + (-1)^{i+j+k-1}  n_l''(\omega_+) \frac{2 l_{{\rm ep},l}(\omega_+)}{\lambda_{{\rm N}l}} \calzzeta_{lj}(\omega_+)_{21} \calzzeta_{lk}(-\omega_-)_{11} \big\}.
\end{split}
\label{eq:phonon_appendix_helper}
\end{align}
\end{widetext}
The factors $n'$ and $n''$ in Eq.~\eqref{eq:phonon_appendix_helper} result from the integral in Eq.~\eqref{eq:phonon_source_integral} with $d_{\rm N} \gg \lambda_{\rm N}$ ({\it c.f.} Eq.~\eqref{eq:jouleheatingfi}),
\begin{align}
\begin{split}
    n_l'(\omega) &= \frac{\lambda_{{\rm N}l} l_{{\rm ep},l}(\omega) - \lambda_{{\rm N}l}^2 \tanh{(d_{\rm N}/l_{{\rm ep},l}(\omega))}}{l_{{\rm ep},l}^2(\omega)- \lambda_{{\rm N}l}^2}, \\
    n_l''(\omega) &= \frac{2 \lambda_{{\rm N}l} l_{{\rm ep},l}(\omega) - \lambda_{{\rm N}l}^2 \tanh{(d_{\rm N}/l_{{\rm ep},l}(\omega))}}{4 l_{{\rm ep},l}^2(\omega)- \lambda_{{\rm N}l}^2}.
\end{split}
\end{align}

\subsection*{Interfacial contribution}
To obtain the nonlinear response relation (\ref{eq:nonlin_interfacesources}), we expand Eq.~\eqref{eq:ispin} to second order in the potential differences and take the limit $\hbar \omega_0/k_{\rm B} T \to 0$. We find that the response matrices in Eq.~\eqref{eq:nonlin_interfacesources} read
\begin{align}
  {\cal A}_{1 \beta \gamma} =&\,
  \frac{e}{k_{\rm B} T} \frac{1}{\zeta(3/2)}
  \begin{pmatrix}
  2 \zeta (1/2) & -2 \zeta (3/2) \\
     -6 \zeta (3/2) & -15 \zeta (5/2)
  \end{pmatrix}
  {\cal N},
  \nonumber \\
  {\cal A}_{2 \beta \gamma} =&\,
  \frac{e}{k_{\rm B} T} \frac{1}{2 \zeta(3/2)}
  \begin{pmatrix}
  12 \zeta (3/2) & 30 \zeta (5/2) \\
     50 \zeta (5/2) & 175 \zeta (7/2)
  \end{pmatrix}
  {\cal N},
  \nonumber \\
\begin{split}
  {\cal B}_{i, 1 \beta \gamma} =&\ 
  \frac{e}{k_{\rm B} T} \frac{2}{\zeta(3/2)} \frac{4 \pi^{3/2} s}{3 k_{\rm T}^3 {\rm Re}\, g_{\uparrow\downarrow i}} \\
  &\, \times
  {\cal N}
  \begin{pmatrix}
  4 \zeta(1/2) & 6 \zeta(3/2) \\ 6 \zeta(3/2) & 15 \zeta(5/2)
  \end{pmatrix}
  {\cal N}, 
\end{split} \\
\begin{split}
  {\cal B}_{i, 2 \beta \gamma} =&\, 
  -\frac{e}{k_{\rm B} T} \frac{5}{\zeta(3/2)} \frac{4 \pi^{3/2} s}{3 k_{\rm T}^3 {\rm Re}\, g_{\uparrow\downarrow i}} \\
  &\, \times
  {\cal N}
  \begin{pmatrix}
  4 \zeta(3/2) & 10 \zeta(5/2) \\ 10 \zeta(5/2) & 35 \zeta(7/2)
  \end{pmatrix}
  {\cal N},
\end{split} \nonumber
\end{align}
where we abbreviated
\begin{equation}
  {\cal N} =
  \begin{pmatrix}
  4 \zeta (3/2) & 10 \zeta (3/2) \\
     10 \zeta (3/2) & 35 \zeta (5/2)
  \end{pmatrix}^{-1}.
\end{equation}

Since we already know the linear response of the spin accumulation to an applied electric field from Eq.~\eqref{eq:linresponse},
\begin{align}
  {\cal U}_{{\rm e}i} (\omega)
  =&\, \sum_{j=1}^2
  (-1)^{j-1} \calzzeta_{ij}(\omega)
  \begin{pmatrix}
  2 \lambda_{{\rm N}j} \theta_{{\rm SH}j} E_j m_y \\ 0 \end{pmatrix}
\label{eq:uresponsefield}
\end{align}
and the linear relation between voltages and currents from Eqs.~\eqref{eq:nmspinrelation} and \eqref{eq:nmheatrelation},
\begin{align}
    {\cal U}_{{\rm e}i} (\omega)
    =&\, (-1)^{i-1} {\cal Z}_{{\rm N}i}(\omega)
  \nonumber \\ &\, \mbox{} \times
  \left[{\cal I}_i(\omega) + 
    \begin{pmatrix}
    \theta_{{\rm SH}i} \sigma_{{\rm N}i} E_i \ve_y \\ 0 \end{pmatrix} \right],
\label{eq:nmspinheatrelation}
\end{align}
we can also express the currents in response to the applied field as
\begin{align}
\begin{split}
  {\cal I}_i(\omega) =&\, -
  \sum_{j=1}^2
  (-1)^{i+j}  {\cal Z}_{{\rm N}i}^{-1}(\omega) \tilde \calzzeta_{ij}(\omega)
  \\ &\, \times
  \begin{pmatrix}
  2 \lambda_{{\rm N}j} \theta_{{\rm SH}j} E_j(\omega) m_y \\ 0 \end{pmatrix}.
\end{split}
\label{eq:iresponsefield}
\end{align}

Inserting Eqs.~\eqref{eq:uresponsefield} and \eqref{eq:iresponsefield} into Eq.~\eqref{eq:nonlin_interfacesources} we obtain the source current
\begin{widetext}
\begin{align}
\begin{split}
  \delta {\cal I}_{i\alpha}^{\rm in} =&\ 
  4 |\vm_{\rm eq} \cdot \ve_y|^2 \sum_{j,k=1}^2 \int \frac{{\rm d}\omega}{2 \pi}
  \lambda_{{\rm N}j} \lambda_{{\rm N}k}
  \theta_{{\rm SH}j} \theta_{{\rm SH}k}
  E_j(\omega_+) E_k(\omega_-)^*
  \\ &\,
  \times (-1)^{j+k} \sum_{\beta, \gamma}
  [{\cal Z}_{{\rm N}i}(\omega_-)^{*-1} 
  \tilde \calzzeta_{ik}(\omega_-)^*]_{\gamma 1}
  \{
  [{\cal Z}_{{\rm N}i}(\omega_+)^{-1} 
  \tilde \calzzeta_{ij}(\omega_+)]_{\beta 1}
  {\cal B}_{i,\alpha\beta\gamma}
  - \calzzeta_{ij}(\omega_+)_{\beta 1}
  {\cal A}_{\alpha\beta\gamma} \}.
\end{split}
\label{eq:dimensionlessNonlinearInterfaceSources}
\end{align}
\end{widetext}

The average charge current from the nonlinear interface relation reads, using Eqs.~\eqref{eq:ix}, \eqref{eq:iy}, \eqref{eq:linresponse}, and \eqref{eq:dimensionlessNonlinearInterfaceSources},
\begin{align}
\begin{split}
  \delta \bar i_{i}^{x}(\omega) =&\, -(-1)^{i-1} \theta_{{\rm SH}i}
     \frac{  \sigma_{{\rm N}i}}{2 d_{{\rm N}i}}
     \sum_{l=1}^{2} (-1)^{l-1} \\
     &\, \times \sum_{\alpha} \left[\tilde \calzzeta_{il}(\Omega) {\cal Z}_{{\rm FN}l\parallel} \right]_{1 \alpha}
   \delta {\cal I}_{l\alpha}^{\rm in}(\Omega) m_y.
\end{split}
\end{align}
Finally, we identify the response coefficients as Eq.~\eqref{eq:interface} of the main text.

\subsection*{Spin-torque contribution}
In the linear response theory of Sec.~\ref{sec:linearresponse}, we have used the equilibrium magnetization direction $\vm_{\rm eq}$ as the reference direction to define longitudinal and transverse spin currents. For the response bilinear in the driving field, the deviations of $\vm$ from the equilibrium direction $\vm_{\rm eq}$ must be taken into account when defining the longitudinal and transverse contributions. Hereto, we write the magnetization at the interface with N$i$ as in Eq.\ (\ref{eq:magbasis}) of the main text.

To define longitudinal and transverse spin transport with respect to the instantaneous magnetization direction $\vm_{i}(t)$, we define the time-dependent complex basis vector $\tilde \ve_{{\perp}i}(t)$ such that it satisfies the time-dependent version of Eq.~\eqref{eq:basiscrossproduct},
\begin{equation}
  \tilde \ve_{{\perp}i}(t) \times \vm_i(t) = i \tilde \ve_{{\perp}i}(t).
\end{equation}
To first order in the magnetization amplitude $m_{\perp i}(t)$ one then has
\begin{align}
\begin{split}
  \vm_i(t) =&\ \vm_{\rm eq} +
  m_{\perp i}(t) \ve_{\perp} +
  m_{\perp i}(t)^* \ve_{\perp}^*, \\
  \tilde \ve_{\perp i}(t) =&\ \ve_{\perp} - m_{\perp i}^*(t) \vm_{\rm eq}.
  \label{eq:etilde}
\end{split}
\end{align}
The longitudinal and transverse spin currents through the F$|$N interfaces depend on the longitudinal and transverse components of the spin accumulation in N, which must be calculated relative to the basis vectors $\vm_i(t)$ and $\tilde \ve_{i\perp}(t)$, respectively. From Eq.~\eqref{eq:etilde} we find
\begin{align}
\begin{split}
  \tilde u_{{\rm s}i\parallel}(t) =&\,
  u_{\parallel i}(t)
  + 2 \mbox{Re}\,  m_{\perp i}^*(t) u_{{\rm s}i \perp}(t), \\
  \tilde u_{{\rm s}i\perp}(t) =&\,
  u_{\perp i}(t) - m_{\perp i}(t) u_{{\rm s}i \parallel}(t),
\end{split}
\end{align}
where $u_{{\rm s}\parallel}(t)$ and $u_{{\rm s}\perp}(t)$ are the longitudinal and transverse components of $\vus(t)$ defined with respect to the equilibrium magnetization direction $\vm_{\rm eq}$. Longitudinal and transverse components $\tilde i_{{\rm s}i\parallel}(t)$ and $\tilde i_{{\rm s}i\perp}(t)$ of the spin currents at the interfaces, defined with respect to the $\vm_i(t)$-dependent basis vectors at the interfaces, can then be calculated from the linear response theory, replacing $u_{{\rm s}i\parallel}(t)$ and $u_{{\rm s}i\perp}(t)$ by $\tilde u_{{\rm s}i\parallel}(t)$ and $\tilde u_{{\rm s}i\perp}(t)$, respectively,
\begin{align}
  \label{eq:fnn2}
  \tilde i_{{\rm s}i\perp}(t) =&\ -(-1)^{i-1} g_{\uparrow\downarrow i}
  [\tilde u_{{\rm s}i \perp}(t) + i (\hbar / e) \dot m_{\perp i}(t)], \\
  \tilde {\cal I}_{i}(t) =&\,
  -(-1)^{i-1} \mathcal{Z}_{{\rm FN}i\parallel}^{-1}
  [\tilde {\cal U}_{{\rm e}i}(t) - {\cal U}_{{\rm m}i}(t)].
  \label{eq:fnn1}
\end{align}
Here we used the two-component vector notation with $\tilde {\cal I}_{i}(t) = (\tilde i_{si \parallel}(t), i_{\rm Qi}(t))^{\rm T}$ and $\tilde {\cal U}_{{\rm e}i}(t) = (\tilde u_{{\rm s}i\parallel}(t),u_{{\rm eQ}i}(t))^{\rm T}$. We express the relation
\begin{align}
  \vi_{{\rm s}i}(t) = \tilde i_{{\rm s} i \parallel}(t) \vm_{i}(t) +
  2 \mbox{Re}\,
  \tilde i_{{\rm s}i\perp}(t) \tilde \ve_{\perp i}(t)
\end{align}
in terms of its longitudinal and transverse components defined with respect to the time-independent basis vectors $\vm_{\rm eq}$ and $\ve_{\perp}$ and obtain
\begin{align}
\begin{split}
  i_{{\rm s}i\parallel}(t) =&\,
  \tilde i_{{\rm s}i\parallel}(t)
  - 2 \mbox{Re}\, m_{\perp i}^*(t) \tilde i_{{\rm s}i\perp}(t),
  \\
  i_{{\rm s}i\perp}(t) =&\,
  \tilde i_{{\rm s}i\perp}(t) + m_{\perp i}(t) \tilde i_{{\rm s}i \parallel}(t).
  \label{eq:nonlin_is}
\end{split}
\end{align}

These equations can be used as the starting point to calculate the spin currents through the F$|$N interfaces to second order in the driving fields $E_{i}(t)$, $i=1,2$. For longitudinal spin transport and for heat transport across the interface and for the transverse component we thus find Eq.\ (\ref{eq:ff2}) of the main text. Explicitly calculating the source terms $\delta {\cal I}_i^{\rm to}$ and $\delta i_{{\rm s}\perp i}^{\rm to}$ we find
\begin{widetext}
\begin{align}
  \begin{split}
  \delta {\cal I}_i^{\rm to}(\Omega)
  =&\
  |\ve_{\perp} \cdot \ve_y|^2
  \sum_{j,k}
  \int \frac{\mathrm{d}\omega}{2 \pi}
   E_j(\omega_+) E_k(\omega_-)^*
  \theta_{{\rm SH}j} \theta_{{\rm SH}k} \lambda_{{\rm N}k}
  \sigma_{{\rm N}j} (-1)^{i+j+k-1}
  \\ &\, \mbox{} \times
  \left[
      [\eta_{ij\perp}(\omega_+) \zzeta_{ik\perp}(\omega_-)^* +
        \eta_{ij\perp}(-\omega_+)^* \zzeta_{ik\perp}(-\omega_-) ]
   {\cal Z}_{{\rm FN}i\parallel}^{-1}
  \begin{pmatrix} 1 \\ 0 \end{pmatrix}
  \right. \\ &\ \ \ \ \left.\ \mbox{}
  - Z_{{\rm N}i}^{-1} [\eta_{ij\perp}(\omega_+) \tilde \zzeta_{ik\perp}(\omega_-)^* +
    \eta_{ij\perp}(-\omega_+)^* \tilde \zzeta_{ik\perp}(-\omega_-) ]
   \begin{pmatrix}
  1 \\ 0 \end{pmatrix}     
  \right], \label{eq:djzlong2}
  \end{split} \\
  \begin{split}
  \delta i_{{\rm s}i\perp}^{\rm to}(\Omega) =&\,
  -
  (\ve_{\perp}^* \cdot \ve_y)(\vm_{\rm eq} \cdot \ve_y)
  \sum_{j,k}
  \int \frac{\mathrm{d}\omega}{2 \pi}
   E_j(\omega_+) E_j(\omega_-)^*
  \theta_{{\rm SH}j} \theta_{{\rm SH}k} \lambda_{{\rm N}k}
  \sigma_{{\rm N}j}
  (-1)^{i+j+k-1}
  \\ &\, \mbox{} \times \eta_{ij\perp}(\omega_+)
  \left[ g_{\uparrow\downarrow i} 
  [\calzzeta_{ik}(-\omega_-)]_{11} -
  \left[{\cal Z}_{{\rm N}i}^{-1}(-\omega_-) \tilde \calzzeta_{ik}(-\omega_-) \right]_{11} \right],
   \label{eq:djzperp2}
   \end{split}
\end{align}
\end{widetext}
where we used a similar expression to Eq.~\eqref{eq:iresponsefield} for the linear-in-$E$ transverse spin current,
\begin{align}
\begin{split}
  i_{{\rm s}i \perp}(\omega) =&\, - 2
  \sum_{k=1}^2
  (-1)^{i+k}  Z_{{\rm N}i}^{-1}(\omega) \tilde \zzeta_{ik\perp}(\omega)
  \\ &\, \times
  \lambda_{{\rm N}k} \theta_{{\rm SH}k} E_k(\omega) m_y.
\end{split}
\end{align}

The bilinear currents in Eqs.~\eqref{eq:djzlong2} and \eqref{eq:djzperp2} are related to the source voltages in the linear response Eq.~\eqref{eq:linresponse} through a similar variable change as discussed in Eq.~\eqref{eq:tildeu_substitution} of Sec.~\ref{sec:interface_nonlin}.

\section{Order-of-magnitude estimates}
\label{app:estimates}

To illustrate the scaling of the bilinear response with the device and material parameters and with frequency, we now present order-of-magnitude estimates for the bilinear response coefficients. These estimates are based on estimates for the dimensionless coefficients $\zzeta_{ij\perp}(\omega)$ and $\calzzeta_{ij}(\omega)$ from the linear response theory of Ref.~\onlinecite{Franke2025-lin}, which we do not repeat here. 

As in Ref.\ \onlinecite{Franke2025-lin}, we assume $\omega \tau_{\rm ep} \lesssim 1$, so that the thermal relaxation length $l_{\rm ep}$ is approximately frequency-independent. We also assume that the thickness $d_{\rm F}$ of the ferromagnetic layer is much larger than the characteristic magnon relaxation length $l_{\mu}$ and $l_{T}$, which are the eigenvalues of $\Lambda(\omega)$, see Eq.\ (\ref{eq:Lambda}). In the low-frequency limit, the larger of these two lengths, $l_{\mu}$, predominantly relaxes the magnon chemical potential $\mu_{\rm m}$, whereas the smaller of the two, $l_{T}$, relaxes the difference $\Delta T_{\rm m}$ of the magnon temperature and the bath temperature \cite{Cornelissen2016-wy}. We also use $l_{\mu}$ to denote the larger of the two relaxation length in the high-frequency regime, where an interpretation as a ``chemical potential relaxation length'' and a ``temperature relaxation length'' is no longer appropriate. We further abbreviate
\begin{align}
  z_{{\rm FN},i} =&\, \frac{s}{k_{\rm T}^3 \mbox{Re}\, g_{\uparrow\downarrow,i}}, \\
  z_{\rm F}(\Omega) =&\, \frac{\hbar^2 \Dex}{2e^2 k_{\rm B} T \sqrt{\tau_{\rm m} \max(1/\tau_{\rm rel},\Omega)}},
\end{align}
where $\tau_{\rm m}$ is the total momentum relaxation time and $\tau_{\rm rel} = \min(\tau_{{\rm m},{\rm rel}},\tau_{{\rm mp},{\rm rel}})$ describes the faster of the two spin-non-conserving relaxation processes in F, and set
\begin{equation}
  z_i(\Omega) = \max(z_{{\rm FN},i},z_{\rm F}(\Omega)).
\end{equation}

{}From the theory of Sec.~\ref{sec:joule_nonlin} and making use of the explicit expressions for the dimensionless linear response coefficients $\zzeta_{ij\perp}(\omega)$ and $\calzzeta_{ij}(\omega)$, see App.\ \ref{sec:impedances},we obtain the order-of-magnitude estimates
\begin{align}
    v_{111}^{\rm Jo} &= w_{111}^{\rm Jo} \sim -  \theta_{\rm SH1} \frac{k_{\rm B} \tau_{{\rm ep},1}}{e C_{{\rm e}1}} \frac{\lambda_{{\rm N}1}}{z_{1}(\Omega)}, \\
    v_{211}^{\rm Jo} &= w_{211}^{\rm Jo}  \sim -  \theta_{\rm SH2} \frac{k_{\rm B} \tau_{{\rm ep},1}}{e C_{{\rm e}1}} \frac{\sigma_{{\rm N}1}}{\sigma_{{\rm N}2}}\frac{\lambda_{{\rm N}2} z_{\rm F}(\Omega)}{z_{1}(\Omega) z_{2}(\Omega)} e^{-d_{\rm F}/l_{\mu}} ,
    \label{eq:low_freq_approx_vijk}
\end{align}
for the local and nonlocal Joule heating contributions to the bilinear response. For the local response, these expressions apply equally to the {\it dc} response at $\Omega = 0$ and to the {\it ac} response at frequency $\Omega = 2 \omega$. For the nonlocal response, these expressions apply for the {\it dc} and {\it ac} response at low driving frequency $\omega$, as well as for the {\it dc} response at large driving frequency. For the {\it ac} response at $\Omega = 2 \omega$ for large driving frequency $\omega$, there is a sign change with respect to the low-frequency limit and a faster decay with the thickness $d_{\rm F}$,
\begin{align}
    v_{211}^{\rm Jo, ac} = w_{211}^{\rm Jo, ac} \sim &\ \theta_{\rm SH2} \frac{k_{\rm B} \tau_{{\rm ep},1}}{e C_{{\rm e}1}} \frac{\sigma_{{\rm N}1}}{\sigma_{{\rm N}2}}\frac{\lambda_{{\rm N}2} z_{\rm F}(\Omega)}{z_{1}(\Omega) z_{2}(\Omega)} e^{-d_{\rm F}/l_{\mu}(\Omega)} ,
    \label{eq:low_freq_approx_vijk_ac}
\end{align}
where $l_{\mu}(\omega) \propto \omega^{-1/2}$ is the larger of the two relaxation lengths in F.

The response coefficient for the local phonon-mediated USMR reads
\begin{equation}
    v_{111}^{\rm ph} \sim \theta_{{\rm SH}1} \frac{\tau_{{\rm ep},1}}{C_{{\rm e}1}} \frac{k_{\rm B}}{e} g_1(\omega_+) \alpha_{{\rm T},1} T \frac{\lambda_{{\rm N}1}}{z_1(\omega_+)}
\end{equation}
and strongly depends on electron-phonon scattering as well as relaxation of phonon energy to a substrate or bath (see definition of $g_i(\omega)$ in Eq.~\eqref{eq:gi}). We note that, in contrast to Joule heating, there is no nonlocal response linear in the spin-Hall angle and only a negligible Hall response from dissipative spin currents (see App.~\ref{app:bilinear}).

In the low-frequency limit, the {\it dc} and {\it ac} contributions to the interfacial USMR to the bilinear response scale as
\begin{align}
\begin{split}
    v_{111}^{\rm in} = &\ w_{111}^{\rm in} = -r_{111}^{\rm in} \\
    \sim &\, - \theta_{{\rm SH}1}^3 \frac{\lambda_{{\rm N}1}^2}{i_{{\rm FN}1}}
    \frac{\lambda_{{\rm N}1}}{\sigma_{{\rm N}1}}
    \frac{1}{z_1 z_{{\rm FN},1}},
  \label{eq:r111estimate}
\end{split} \\
\begin{split}
\label{eq:low_freq_approx_rijk}
  v_{211}^{\rm in} = &\ w_{211}^{\rm in} = -r_{211}^{\rm in} \\
  \sim &\, - \theta_{{\rm SH}1}^2 \theta_{{\rm SH}2} \frac{\lambda_{{\rm N}1}^2}{i_{{\rm FN}2}}
    \frac{\lambda_{{\rm N}2}}{\sigma_{{\rm N}2}}
    \frac{z_{\rm F}^2}{z_1^2 z_2^2} e^{-2 d_{\rm F}/l_{\mu}}
\end{split}
\end{align}
with the caveat that the numerical prefactor, which is not included in Eq.~\eqref{eq:r111estimate}, is $\sim 10^{-2}$. Here, the characteristic interfacial current density
\begin{equation}
    i_{{\rm FN}i} = \frac{3 \mathrm{Re} (g_{\uparrow\downarrow i})} {4 \pi^{3/2} e s} k_{\rm T}^3 k_{\rm B} T.
\label{eq:jfn}
\end{equation}
At finite driving frequencies $\omega$, the local {\it dc} and {\it ac} interfacial USMR response stays of the order of magnitude given in the zero-frequency estimate, Eq.~\eqref{eq:r111estimate}, and the estimate (\ref{eq:low_freq_approx_rijk}) for the nonlocal response also applies to the {\it dc} response at finite driving frequency. However, the nonlocal {\it ac} response is exponentially suppressed with $d_{\rm F} / l_{\mu}(\Omega)$, where $l_{\mu}(\Omega)$ is the largest eigenvalue of $\Lambda(\Omega)$. As \citet{Sterk2019-mt}, we find that for YIG and Pt, the interfacial USMR effect is negligible compared to the spin-Seebeck effect from Joule heating for the local bilinear response, which has a similar experimental signature. In a nonlocal measurement, the interfacial USMR contribution may be visible because of its characteristic magnetization dependence $\propto m_{y/x} m_y^2$.

The low-frequency approximation for the spin-torque contribution reads
\begin{align}
  w_{111}^{\rm to}  \sim&\ \theta_{{\rm SH}1}^3 \frac{\lambda^2_{{\rm N}1}}{i_{\rm F}} \frac{\lambda_{{\rm N}1}}{\sigma_{{\rm N}1}} \mathrm{Re} g_{\uparrow\downarrow 1}^2, \\
  \begin{split}
  w_{211}^{\rm to}  \sim&\ \theta_{{\rm SH}1}^2 \theta_{{\rm SH}2} \frac{\lambda^2_{{\rm N}1} }{i_{\rm F}} \frac{\lambda^2_{{\rm N}2}}{\sigma^2_{{\rm N}2}} \\
  &\, \times \mathrm{Im} (g_{\uparrow\downarrow 1} g_{\uparrow\downarrow 2}) \frac{z_{\rm F}}{z_1 z_2} e^{-d_{\rm F}/l_{\mu}}, \end{split} \\
  r_{211}^{\rm to}  \sim&\ \theta_{{\rm SH}1}^2 \theta_{{\rm SH}2} \frac{\lambda_{{\rm N}1}^2}{i_{\rm F}}\frac{\lambda_{{\rm N}2}}{\sigma_{{\rm N}2}} \mathrm{Im} g_{\uparrow\downarrow 1} \frac{z_{\rm F}}{z_1 z_2} e^{-d_{\rm F}/l_{\mu}}, \\
  \begin{split}
  t_{211}^{\rm to} \sim&\ \theta_{{\rm SH}1}^2 \theta_{{\rm SH}2} \frac{\lambda^2_{{\rm N}1} }{i_{\rm F}} \frac{\lambda^2_{{\rm N}2}}{\sigma^2_{{\rm N}2}} \\
  &\, \times \mathrm{Re} (g_{\uparrow\downarrow 1} g_{\uparrow\downarrow 2}) \frac{z_{\rm F}}{z_1 z_2} e^{-d_{\rm F}/l_{\mu}}, \end{split}
\end{align}
with
\begin{equation}
  i_{\rm F} = e s \sqrt{\Dex \omega_0}.
  \label{eq:iF}
\end{equation}
The local bilinear response coefficients $r_{111}^{\rm to}$ and $t_{111}^{\rm to}$ have the same order of magnitude as $w_{111}^{\rm to}$.
The bilinear spin-torque response at large driving frequency $\omega$ is dominated by resonances at $\omega = \omega_n$ and $\omega = \omega_n/2$, see Eq.~\eqref{eq:resonancefrequencies}.
The peak value of the nonlinear spin-torque response at the resonances $\omega = \omega_n$ scales as
\begin{align}
  |w_{111}^{\rm to, dc} |_{\rm peak} \sim&\
  |w_{111}^{\rm to, ac} |_{\rm peak} \nonumber \\ \sim&\
  \theta_{{\rm SH}1}^3 
  \frac{\lambda_{{\rm N}1}^3}{\sigma_{{\rm N}1}} 
  \frac{e}{\hbar \omega_n} \mathrm{Re} g_{\uparrow\downarrow 1},
\end{align}
with the same estimate for the peaks of the response coefficients $r_{111}^{\rm to}$ and $t_{111}^{\rm to}$. Similarly, we find for the nonlocal {\it dc} spin-torque contribution at the resonance peaks
\begin{align}
  |r_{211}^{\rm to, dc}|_{\rm peak} \sim&\
  \theta_{{\rm SH}1}^2 \theta_{{\rm SH}2}
  \frac{\lambda_{{\rm N}1}^2 \lambda_{{\rm N}2}}{\sigma_{{\rm N}2}} \frac{e}{\hbar \omega_n} \mathrm{Re} g_{\uparrow\downarrow 1},
\end{align}
whereas $w_{211}^{\rm to, dc}$ and $t_{211}^{\rm to, dc}$ scale with the exponent $- d_{\rm F} / l_{\mu}(\Omega)$ and are therefore several orders of magnitude smaller than $r_{211}^{\rm to, dc}$. The nonlocal {\it ac} resonant spin-torque response for all $w_{211}^{\rm to}$, $r_{211}^{\rm to}$, and $t_{211}^{\rm to}$ is of a similar magnitude as $|r_{211}^{{\rm to, dc}}|_{\rm peak}$, although the individual peak heights vary with $\cos{(k(\omega)d_{\rm F}})$.

\end{appendix}

\typeout{}
\bibliographystyle{apsrev4-1}
\bibliography{main}

\end{document}